# Convergence of the Discrete Dipole Approximation.

# Part 1: Theoretical Analysis.

**Maxim A. Yurkin**

Faculty of Science, Section Computational Science, of the University of Amsterdam,

Kruislaan 403, 1098 SJ, Amsterdam, The Netherlands

and

Institute of Chemical Kinetics and Combustion, Siberian Branch of the Russian Academy of

Sciences, Institutskaya 3, Novosibirsk 630090 Russia

myurkin@science.uva.nl

**Valeri P. Maltsev**

Institute of Chemical Kinetics and Combustion, Siberian Branch of the Russian Academy of

Sciences, Institutskaya 3, Novosibirsk 630090 Russia

and

Novosibirsk State University, Pirogova Str. 2, 630090, Novosibirsk, Russia

**Alfons G. Hoekstra**

Faculty of Science, Section Computational Science, of the University of Amsterdam,

Kruislaan 403, 1098 SJ, Amsterdam, The Netherlands

alfons@sciene.uva.nl

## Abstract

We performed a rigorous theoretical convergence analysis of the Discrete Dipole Approximation (DDA). We prove that errors in any measured quantity are bounded by a sum of a linear and quadratic term in the size of a dipole *d*, when the latter is in the range of DDA applicability. Moreover, the linear term is significantly smaller for cubically than for non-cubically shaped scatterers. Therefore, for small *d* errors for cubically shaped particles are much smaller than for non-cubically shaped. The relative importance of the linear term decreases with increasing size, hence convergence of DDA for large enough scatterers is quadratic in the common range of *d*. Extensive numerical simulations were carried out for a wide range of *d*. Finally we discuss a number of new developments in DDA and their consequences for convergence.

## 1. Introduction

The Discrete Dipole Approximation (DDA) is a well-known method to solve the light scattering problem for arbitrary shaped particles. Since its introduction by Purcell and Pennypacker[1] it has been improved constantly. The formulation of DDA summarized by Draine and Flatau[2] more than 10 years ago is still most widely used for many applications,[3] partly due to the publicly available high-quality and user-friendly code DDSCAT.[4] Although modern improvements of DDA (as discussed in detail in Section 2.F) exist, they are still in the research stage because they are not widely used in real applications.

DDA directly discretizes the volume of the scatterer and hence is applicable to arbitrary shaped particles. However, the drawback of this discretization is the extreme computational complexity of DDA, although it is significantly decreased by advanced numerical techniques.[2,5] That is why the usual application strategy for DDA is "single computation", where a discretization is chosen based on available computational resources and some empirical estimates of the expected errors.[3,4] These error estimates are based on a limited number of benchmark calculations[3] and hence are external to the light scattering problem under investigation. Such error estimates have evident drawbacks, however no better alternative is available. Some results of analytical analysis of errors in computational electromagnetics are known, e.g. [6,7], however they typically consider the surface integral equations. To the best of our knowledge, such analysis has not been done for volume integral equations (such as DDA).

Usually errors in DDA are studied as a function of the size parameter of the scatterer $x$ (at a constant or few different total numbers of dipoles $N$), e.g. [2,8]. Only a small number of papers directly present errors versus discretization parameter (e.g. $d$ – the size of a single dipole).[9–17] The range of $d$ typically studied in those papers is limited to a 5 times difference between minimum and maximum values, with the exception of two papers[11,12] where it is 15 times. Those plots of errors versus discretization parameter are always used to illustrate the performance of a new DDA formulation and compare it with others. No conclusions about the convergence properties of DDA, as a function of $d$, have been made based on these plots. To our knowledge, no theoretical analysis of DDA convergence has been performed, but only a few limited empirical studies have appeared in the literature.

In this paper we perform a theoretical analysis of DDA convergence when refining the discretization (Section 2). We derive rigorous theoretical bounds on the error in any measured quantity for any scatterer. In Section 3 we present extensive numerical results of DDA computations for 5 different scatterers using many different discretizations. These results are discussed in Section 4 to support conclusions of the theoretical analysis. We formulate the conclusions of the paper in Section 5. In a follow-up paper[18] (which from now on we refer to as Paper 2) the theoretical convergence results are used for an extrapolation technique to increase the accuracy of DDA computations.

## 2. Theoretical analysis

In this section we analyze theoretically the errors of DDA computations. We formulate the volume integral equation for the internal electric field and its operator counterpart in Section 2.A and its discretization in Section 2.B. Section 2.C contains integral and discretized formulae for measured quantities that are the final goal of any light scattering simulation. We derive the main results in Section 2.D, where we consider errors of the traditional DDA formulation[2] without shape errors, which are considered separately in Section 2.E. Finally in Section 2.F we discuss some recent DDA improvements from the viewpoint of our convergence theory.

### *A. Integral Equation*

Throughout this paper we assume the $\exp(-\mathrm{i}\omega t)$ time dependence of all fields. The scatterer is assumed dielectric but not magnetic (magnetic permittivity $\mu = 1$ ), and the electric permittivity is assumed isotropic (non-isotropic permittivity will significantly complicate the derivations but will not principally change the main conclusion of Section 2 – Eqs. (70) and (87)).

The general form of the integral equation governing the electric field inside the dielectric scatterer is the following:[19, 20]

$$\mathbf{E}(\mathbf{r}) = \mathbf{E}^{inc}(\mathbf{r}) + \int_{V\setminus V_0} d^3r' \overline{\mathbf{G}}(\mathbf{r},\mathbf{r}')\chi(\mathbf{r}')\mathbf{E}(\mathbf{r}') + \mathbf{M}(V_0,\mathbf{r}) - \overline{\mathbf{L}}(\partial V_0,\mathbf{r})\chi(\mathbf{r})\mathbf{E}(\mathbf{r}), \tag{1}$$

where $\mathbf{E}^{inc}(\mathbf{r})$, $\mathbf{E}(\mathbf{r})$ are the incident and total electric field at location $\mathbf{r}$; $\chi(\mathbf{r}) = (\varepsilon(\mathbf{r}) - 1)/4\pi$ is the susceptibility of the medium at point $\mathbf{r}$ ($\varepsilon(\mathbf{r})$ – relative permittivity). $V$ is the volume of the particle (more general – the volume, which contains all points where the susceptibility is not zero), $V_0$ is a smaller volume such that $V_0 \subset V$, $\mathbf{r} \in V_0 \setminus \partial V_0$. $\overline{\mathbf{G}}(\mathbf{r},\mathbf{r}')$ is the free space dyadic Green's function, defined as

$$\overline{\mathbf{G}}(\mathbf{r},\mathbf{r}') = \left(k^2\overline{\mathbf{I}} + \hat{\nabla}\hat{\nabla}\right)g(R) = g(R)\left[k^2\left(\overline{\mathbf{I}} - \frac{\hat{R}\hat{R}}{R^2}\right) - \frac{1-\mathrm{i}kR}{R^2}\left(\overline{\mathbf{I}} - 3\frac{\hat{R}\hat{R}}{R^2}\right)\right], \tag{2}$$

where $\overline{\mathbf{I}}$ is the identity dyadic, $k = \omega/c$ – free space wave vector, $\mathbf{R} = \mathbf{r} - \mathbf{r}'$, $R = |\mathbf{R}|$, and $\hat{R}\hat{R}$ is a dyadic defined as $\hat{R}\hat{R}_{\mu\nu} = R_\mu R_\nu$ ($\mu$, $\nu$ are Cartesian components of the vector or tensor), and $g(R)$ is the scalar Green's function

$$g(R) = \frac{\exp(\mathrm{i}kR)}{R}. \tag{3}$$

$\mathbf{M}$ is the following integral associated with the finiteness of the exclusion volume $V_0$

$$\mathbf{M}(V_0,\mathbf{r}) = \int_{V_0} d^3r' \left(\overline{\mathbf{G}}(\mathbf{r},\mathbf{r}')\chi(\mathbf{r}')\mathbf{E}(\mathbf{r}') - \overline{\mathbf{G}}^s(\mathbf{r},\mathbf{r}')\chi(\mathbf{r})\mathbf{E}(\mathbf{r})\right), \tag{4}$$

where $\overline{\mathbf{G}}^s(\mathbf{r},\mathbf{r}')$ is the static limit ($k \to 0$) of $\overline{\mathbf{G}}(\mathbf{r},\mathbf{r}')$:

$$\overline{\mathbf{G}}^s(\mathbf{r},\mathbf{r}') = \hat{\nabla}\hat{\nabla}\frac{1}{R} = -\frac{1}{R^3}\left(\overline{\mathbf{I}} - 3\frac{\hat{R}\hat{R}}{R^2}\right). \tag{5}$$

$\overline{\mathbf{L}}$ is the so-called self-term dyadic:

$$\overline{\mathbf{L}}(\partial V_0,\mathbf{r}) = -\oint_{\partial V_0} d^2r' \frac{\hat{n}'\hat{R}}{R^3}, \tag{6}$$

where $\hat{n}'$ is an external (as viewed from $\mathbf{r}$) normal to the surface $\partial V_0$ at point $\mathbf{r}'$.

Eq. (1) can be rewritten in operator form as follows

$$\tilde{\mathbf{A}} \cdot \tilde{\mathbf{E}} = \tilde{\mathbf{E}}^{inc}, \tag{7}$$

where $\tilde{\mathbf{E}} \in H_1 = L^1(V \to \mathbf{C}^3)$ – functions from $V$ to $\mathbf{C}^3$ that have finite $L^1$-norm, $\tilde{\mathbf{E}}^{inc} \in H_2$ – subspace of $H_1$ containing all functions that satisfy Maxwell equations in free space. $\tilde{\mathbf{A}}$ is a linear operator $: H_1 \to H_2$. Although the Sobolev norm is physically more sound (based on the finiteness of energy of the electric field),[6, 21] we use the $L^1$-norm. A detailed discussion of all assumptions made for the electric field is performed in Section 2.D.

### *B. Discretization*

To solve Eq. (1) numerically a discretization is done in the following way.[20] Let $V = \bigcup_{i=1}^{N} V_i$, $V_i \cap V_j = \emptyset$ for $i \neq j$. $N$ denotes the number of subvolumes (dipoles). Assuming $\mathbf{r} \in V_i$ and choosing $V_0 = V_i$, Eq. (1) can be rewritten as

$$\mathbf{E}(\mathbf{r}) = \mathbf{E}^{inc}(\mathbf{r}) + \sum_{j \neq i} \int_{V_j} d^3 r' \overline{\mathbf{G}}(\mathbf{r}, \mathbf{r}') \chi(\mathbf{r}') \mathbf{E}(\mathbf{r}') + \mathbf{M}(V_i, \mathbf{r}) - \overline{\mathbf{L}}(\partial V_i, \mathbf{r}) \chi(\mathbf{r}) \mathbf{E}(\mathbf{r}). \tag{8}$$

The set of Eq. (8) (for all *i*) is exact. Further one fixed point $\mathbf{r}_i$ inside each $V_i$ (its center) is chosen and $\mathbf{r} = \mathbf{r}_i$ is set.

The usual approximation[20] is considering $\mathbf{E}$ and $\chi$ constant inside each subvolume:

$$\mathbf{E}(\mathbf{r}) = \mathbf{E}(\mathbf{r}_i) = \mathbf{E}_i,\ \chi(\mathbf{r}) = \chi(\mathbf{r}_i) = \chi_i \ \text{ for } \ \mathbf{r} \in V_i. \tag{9}$$

Eq. (8) can then be rewritten as

$$\mathbf{E}_i = \mathbf{E}_i^{inc} + \sum_{j \neq i} \overline{\mathbf{G}}_{ij} V_j \chi_j \mathbf{E}_j + \left( \overline{\mathbf{M}}_i - \overline{\mathbf{L}}_i \right) \chi_i \mathbf{E}_i, \tag{10}$$

where $\mathbf{E}_i^{inc} = \mathbf{E}^{inc}(\mathbf{r}_i)$, $\overline{\mathbf{L}}_i = \overline{\mathbf{L}}(\partial V_i, \mathbf{r}_i)$,

$$\overline{\mathbf{M}}_i = \int_{V_i} d^3r' \left( \overline{\mathbf{G}}(\mathbf{r}_i, \mathbf{r}') - \overline{\mathbf{G}}^s(\mathbf{r}_i, \mathbf{r}') \right), \tag{11}$$

$$\overline{\mathbf{G}}_{ij} = \frac{1}{V_j} \int_{V_j} d^3r' \overline{\mathbf{G}}(\mathbf{r}_i, \mathbf{r}') . \tag{12}$$

A further approximation, which is used in almost all formulations of DDA, is

$$\overline{\mathbf{G}}_{ij}^{(0)} = \overline{\mathbf{G}}(\mathbf{r}_i, \mathbf{r}_j) . \tag{13}$$

This assumption is made implicitly by all formulations that start by replacing the scatterer with a set of point dipoles, as was done originally by Purcell and Pennypacker.[1] For a cubical (as well as spherical) cell $V_i$ with $\mathbf{r}_i$ located at the center of the cell, $\overline{\mathbf{L}}_i$ can be calculated analytically yielding[22]

$$\overline{\mathbf{L}}_i = \frac{4\pi}{3} \overline{\mathbf{I}} . \tag{14}$$

Eq. (10) together with Eqs. (13) and (14) and completely neglecting $\overline{\mathbf{M}}_i$ is equivalent to the original DDA by Purcell and Pennypacker (PP).[1] The diagonal terms in Eq. (10) are then equivalent to the well-known Clausius-Mossotti (CM) polarizability for point dipoles. Modifications introduced by other DDA prescriptions are discussed in Section 2.F.

In matrix notation Eq. (10) reads

$$\overline{\mathbf{A}}^d \mathbf{E}^d = \mathbf{E}^{inc,d} , \tag{15}$$

where $\mathbf{E}^d$, $\mathbf{E}^{inc,d}$ are elements of $\left(\mathbf{C}^3\right)^N$ (vectors of size $N$ where each element is a 3D complex vector) and $\overline{\mathbf{A}}^d$ is a $N \times N$ matrix where each element is a 3×3 tensor. $d$ is the size of one dipole. In operator notation Eq. (8) (for $\mathbf{r} = \mathbf{r}_i$) is as follows

$$\left(\widetilde{\mathbf{A}}\widetilde{\mathbf{E}}\right)(\mathbf{r}_i) = \widetilde{\mathbf{E}}^{inc}(\mathbf{r}_i) = \mathbf{E}_i^{inc,d} , \tag{16}$$

We define the discretization error function as

$$\mathbf{h}_i^d = \left(\widetilde{\mathbf{A}}\widetilde{\mathbf{E}}\right)(\mathbf{r}_i) - \left(\overline{\mathbf{A}}^d \mathbf{E}^{0,d}\right)_i , \tag{17}$$

where $\mathbf{E}^{0,d}$ is the exact field at the centers of the dipoles – $\mathbf{E}_i^{0,d} = \tilde{\mathbf{E}}(\mathbf{r}_i)$, in contrast to $\mathbf{E}^d$ that is only an approximation obtained from solution of Eq. (15) (here we neglect the numerical error that appears from the solution of Eq. (15) itself, which is acceptable if this error is controlled to be much less than other errors). Comparing Eqs. (15) and (17) one can immediately obtain the error in internal fields due to discretization $\delta\mathbf{E}^d$:

$$\delta\mathbf{E}^d = \mathbf{E}^d - \mathbf{E}^{0,d} = -\left(\overline{\mathbf{A}}^d\right)^{-1}\mathbf{h}^d . \tag{18}$$

## C.Measured quantities

After having determined the internal electric fields, scattered fields and cross sections can be calculated. Scattered fields are obtained by taking the limit $r \to \infty$ of the integral in Eq. (1) (see e.g. [23])

$$\mathbf{E}^{sca}(\mathbf{r}) = \frac{\exp(\mathrm{i}kr)}{-\mathrm{i}kr}\mathbf{F}(\mathbf{n}), \tag{19}$$

where $\mathbf{n} = \mathbf{r}/r$ is the unit vector in the scattering direction, and $\mathbf{F}$ is the scattering amplitude:

$$\mathbf{F}(\mathbf{n}) = -\mathrm{i}k^3(\overline{\mathbf{I}} - \hat{n}\hat{n})\sum_i \int_{V_i} d^3r' \exp(-\mathrm{i}k\mathbf{r}' \cdot \mathbf{n})\chi(\mathbf{r}')\mathbf{E}(\mathbf{r}') . \tag{20}$$

All other differential scattering properties, such as the amplitude and Mueller scattering matrices, and asymmetry parameter $\langle\cos\theta\rangle$ can be easily derived from $\mathbf{F}(\mathbf{n})$, calculated for two incident polarizations.[24] We consider an incident polarized plane wave:

$$\mathbf{E}^{inc}(\mathbf{r}) = \mathbf{e}^0 \exp(\mathrm{i}\mathbf{k} \cdot \mathbf{r}), \tag{21}$$

where $\mathbf{k} = k\mathbf{a}$, $\mathbf{a}$ is direction of incidence, and $\left|\mathbf{e}^0\right| = 1$ is assumed. The scattering and extinction cross sections ($C_{sca}$, $C_{ext}$) are derived from the scattering amplitude:[23]

$$C_{sca} = \frac{1}{k^2}\int d\Omega \left|\mathbf{F}(\mathbf{n})\right|^2 , \tag{22}$$

$$C_{ext} = \frac{4\pi}{k^2}\mathrm{Re}\left(\mathbf{F}(\mathbf{a})\cdot\mathbf{e}^{0*}\right), \tag{23}$$

where * denote complex conjugation. The expression for absorption cross section ($C_{abs}$) directly uses the internal fields:[23]

$$C_{abs} = 4\pi k\sum_i\int_{V_i} d^3r'\,\mathrm{Im}\left(\chi(\mathbf{r}')\right)\left|\mathbf{E}(\mathbf{r}')\right|^2, \tag{24}$$

Since only values of the internal field in the centers of dipoles are known, Eqs. (20) and (24) are approximated by (PP)

$$\mathbf{F}(\mathbf{n}) = -\mathrm{i}k^3(\bar{\mathbf{I}} - \hat{n}\hat{n})\sum_i V_i\chi_i\mathbf{E}_i^d\exp(-\mathrm{i}k\mathbf{r}_i\cdot\mathbf{n}), \tag{25}$$

$$C_{abs} = 4\pi k\sum_i V_i\,\mathrm{Im}(\chi_i)\left|\mathbf{E}_i^d\right|^2. \tag{26}$$

Corrections to Eq. (26) are discussed in Section 2.F.

Both Eqs. (20) (for each component) and (24) can be generalized as $\tilde{\phi}\left(\tilde{\mathbf{E}}\right)$ (a functional that is not necessarily linear), which is approximated as:

$$\tilde{\phi}\left(\tilde{\mathbf{E}}\right) = \phi^d\left(\mathbf{E}^d\right) + \delta\phi^d, \tag{27}$$

where $\phi^d\left(\mathbf{E}^d\right)$ corresponds to Eqs. (25) or (26) respectively, and the error $\delta\phi^d$ consists of two parts:

$$\delta\phi^d = \left[\tilde{\phi}\left(\tilde{\mathbf{E}}\right) - \phi^d\left(\mathbf{E}^{0,d}\right)\right] + \left[\phi^d\left(\mathbf{E}^{0,d}\right) - \phi^d\left(\mathbf{E}^d\right)\right]. \tag{28}$$

The first one comes from discretization (similar to Eq. (17)), and the second from errors in the internal fields.

## *D. Error analysis*

In this section we perform error analysis for the PP formulation of DDA. Improvements of DDA are further discussed in Section 2.F.

We assume cubical subvolumes with size $d$. We also assume that the shape of the particle is *exactly* described by these cubical subvolumes (we call this *cubically shaped* scatterer). Moreover, $\chi$ is a smooth function inside $V$ (exact assumptions on $\chi$ are formulated below). An extension of the theory to shapes that do not satisfy these conditions is presented in Section 2.E. If there are several regions with different values of $\chi$ (smooth inside each region), the analysis is still valid but interfaces inside $V$ should be considered the same way as the outer boundary of $V$. We further fix the geometry of the scattering problem and incident field. Therefore we will be interested only in variation of discretization (which is characterized by the single parameter – $d$); for reasons that will become clear in the sequel, we assume that $kd < 2$ (this bound is not limiting since otherwise DDA is generally inapplicable[2]).

We switch to dimensionless parameters by assuming $k = 1$, which is equivalent to measuring all the distances in units of $1/k$. The unit of electric field can be chosen arbitrary but constant. In all further derivations we will use two sets of constants: $\gamma_i$ and $c_i$. $\gamma_1$-$\gamma_{13}$ are basic constants that do not depend on the discretization $d$, but do depend directly upon all other problem parameters – size parameter $x = kR_{eq}$ ($R_{eq}$ – volume-equivalent radius), $m$, shape, and incident field – or some of them. On the contrary, $c_1$-$c_{94}$ are auxiliary values that either are numerical constants or can be derived in terms of constants $\gamma_i$. Although the dependencies of $c_i$ on $\gamma_i$ are not explicitly derived in this paper, one can easily obtain them following the derivations of this section. That is the main motivation for using such vast amount of constants instead of an “order of magnitude” formalism. However, such explicit derivation has limited application because, as we will see further, constants in the final result depend upon almost all basic constants. Qualitative analysis of these dependencies will be performed in the end of this section. It should be noted that the main theoretical results concerning DDA convergence (boundedness of errors by a quadratic function, cf. Eq (70)) can be formulated and applied without consideration of any constants (which is simpler).

However our full derivation enables us to make additional conclusions related to the behavior of specific error terms.

The total number of dipoles used to discretize the scatterer is

$$N = \gamma_1 d^{-3}. \tag{29}$$

We assume that the internal field $\tilde{\mathbf{E}}$ is at least four times differentiable and all these derivatives are bounded inside $V$

$$\begin{gathered} \left|\mathbf{E}(\mathbf{r})\right| \le \gamma_2, \left|\partial_\mu \mathbf{E}(\mathbf{r})\right| \le \gamma_3, \left|\partial_\mu \partial_\nu \mathbf{E}(\mathbf{r})\right| \le \gamma_4, \left|\partial_\mu \partial_\nu \partial_\rho \mathbf{E}(\mathbf{r})\right| \le \gamma_5, \left|\partial_\mu \partial_\nu \partial_\rho \partial_\tau \mathbf{E}(\mathbf{r})\right| \le \gamma_6 \\ \text{for } \mathbf{r} \in V \text{ and } \forall \mu, \nu, \rho, \tau. \end{gathered} \tag{30}$$

This assumption is acceptable since there are no interfaces inside $V$, therefore $\tilde{\mathbf{E}}$ should be a smooth function. $|.|$ denotes the Euclidian ($L^2$) norm, which is used for all 3D objects: vectors and tensors. We use the $L^1$-norm, $\|.\|_1$, for $N$-dimensional vectors and matrices as well as for functions and operators. Eq. (30) immediately implies that $\tilde{\mathbf{E}} \in L^1(V)$. We require that $\chi$ satisfies Eq. (30) with constants $\gamma_7$-$\gamma_{11}$. Further we will state an estimate for the norm of $\overline{\mathbf{G}}(\mathbf{R})$ and its derivatives. One can easily obtain from Eq. (2) that for $R > 1$ $\overline{\mathbf{G}}(\mathbf{R})$ satisfies Eq. (30) (with constants $c_1$-$c_5$), while for $R \le 2$

$$\begin{gathered} \left|\overline{\mathbf{G}}(\mathbf{R})\right| \le c_6 R^{-3}, \left|\partial_\mu \overline{\mathbf{G}}(\mathbf{R})\right| \le c_7 R^{-4}, \left|\partial_\mu \partial_\nu \overline{\mathbf{G}}(\mathbf{R})\right| \le c_8 R^{-5}, \left|\partial_\mu \partial_\nu \partial_\rho \overline{\mathbf{G}}(\mathbf{R})\right| \le c_9 R^{-6}, \\ \left|\partial_\mu \partial_\nu \partial_\rho \partial_\tau \overline{\mathbf{G}}(\mathbf{R})\right| \le c_{10} R^{-7} \text{ for } \forall \mu, \nu, \rho, \tau. \end{gathered} \tag{31}$$

Next we state two auxiliary facts that will be used later. Let $V_c$ be a cube with size $d$ and with its center at the origin and $f(\mathbf{r})$ a four times differentiable function inside $V_c$. Then

$$\left| \frac{1}{d^3} \int_{V_c} d^3 r\, f(\mathbf{r}) - f(\mathbf{0}) \right| \le c_{11} d^2 \max_{\mu\nu, \mathbf{r} \in V_c} \left|\partial_\mu \partial_\nu f(\mathbf{r})\right|, \tag{32}$$

$$\left| \frac{1}{d^3} \int_{V_c} d^3 r\, f(\mathbf{r}) - f(\mathbf{0}) \right| \le \frac{d^2}{24} \left| \left(\nabla^2 f(\mathbf{r})\right)_{\mathbf{r}=\mathbf{0}} \right| + c_{12} d^4 \max_{\mu\nu\rho\tau, \mathbf{r} \in V_c} \left|\partial_\mu \partial_\nu \partial_\rho \partial_\tau f(\mathbf{r})\right|. \tag{33}$$

Eqs. (32) and (33) are the corollary of expanding $f$ into Taylor series. Odd orders of the Taylor expansion vanish because of cubical symmetry.

Our first goal is to estimate $\left\|\mathbf{h}^d\right\|_1$. Starting from Eq. (17) we write $\mathbf{h}_i^d$ as

$$\mathbf{h}_i^d = \sum_{j\neq i}\left( \int_{V_j} d^3r' \overline{\mathbf{G}}(\mathbf{r}_i,\mathbf{r}')\mathbf{P}(\mathbf{r}') - d^3\overline{\mathbf{G}}_{ij}^{(0)}\mathbf{P}_j \right) + \mathbf{M}(V_i,\mathbf{r}_i), \tag{34}$$

where we have introduced the polarization vector for conciseness

$$\mathbf{P}(\mathbf{r}) = \chi(\mathbf{r})\mathbf{E}(\mathbf{r}),\ \mathbf{P}_i = \mathbf{P}(\mathbf{r}_i). \tag{35}$$

It is evident that $\mathbf{P}(\mathbf{r})$ also satisfies Eq. (30) (with constants $c_{13}$-$c_{17}$). We start by estimating $\left|\mathbf{M}(V_i,\mathbf{r}_i)\right|$. Substituting a Taylor expansion of $\mathbf{P}(\mathbf{r})$

$$\mathbf{P}(\mathbf{R}) = \mathbf{P}(\mathbf{0}) + \sum_{\rho} R_\rho \left(\partial_\rho \mathbf{P}\right)(\mathbf{0}) + \frac{1}{2}\sum_{\rho\tau} R_\rho R_\tau \left(\partial_\rho\partial_\tau \mathbf{P}\right)\left(\tilde{\mathbf{r}}(\rho,\tau,\mathbf{R})\right), \tag{36}$$

where $0 \le \tilde{r}_\mu \le R_\mu$, into Eq (4) gives

$$\mathbf{M}(V_i,\mathbf{r}_i) = \int_{V_i} d^3R\left(\overline{\mathbf{G}}(\mathbf{R}) - \overline{\mathbf{G}}^s(\mathbf{R})\right)\mathbf{P}_i + \frac{1}{2}\int_{V_i} d^3R\overline{\mathbf{G}}(\mathbf{R})\sum_{\rho\tau} R_\rho R_\tau \left(\partial_\rho\partial_\tau \mathbf{P}\right)\left(\tilde{\mathbf{r}}(\rho,\tau,\mathbf{R})\right). \tag{37}$$

The norms of these two terms can be estimated as

$$\left| \int_{V_i} d^3R\left(\overline{\mathbf{G}}(\mathbf{R}) - \overline{\mathbf{G}}^s(\mathbf{R})\right)\mathbf{P}_i \right| = \left| \frac{2}{3}\overline{\mathbf{I}}\mathbf{P}_i \int_{V_i} d^3R g(R) \right| \le c_{18} d^2, \tag{38}$$

$$\left| \int_{V_i} d^3R\overline{\mathbf{G}}(\mathbf{R})\sum_{\rho\tau} R_\rho R_\tau \left(\partial_\rho\partial_\tau \mathbf{P}\right)\left(\tilde{\mathbf{r}}(\rho,\tau,\mathbf{R})\right) \right| \le 3c_{15}\int_{V_i} d^3R\left|\overline{\mathbf{G}}(\mathbf{R})\right| R^2 \le c_{19} d^2. \tag{39}$$

Eq. (38) follows directly from the definitions in Eqs. (2), (5). To derive Eq. (39) we used Eq. (31) and the fact that $\sum_{\rho\tau}\left|R_\rho R_\tau\right| \le 3R^2$. Finally, Eqs (37)-(39) lead to

$$\left|\mathbf{M}(V_i,\mathbf{r}_i)\right| \le c_{20} d^2. \tag{40}$$

To estimate the sum in Eq. (34) we consider separately three cases: 1) dipole $j$ lies in a *complete shell* of dipole $i$ (we define the shell below); 2) $j$ lies in a *distant shell* of dipole $i$ – $R_{ij} = \left|\mathbf{r}_j - \mathbf{r}_i\right| > 1$; 3) all $j$ that fall between the first two cases (see Fig. 1). We define the first *shell* ($S_1(i)$) of a cubical dipole as a set of dipoles that touch it (including touching in one point only). The second shell ($S_2(i)$) is a set of dipoles that touch the outer surface of the first shell, and so on. The $l$-th shell ($S_l(i)$) is then a set of all dipoles that lie on the boundary of the cube with size $(2l+1)d$ and center coinciding with the center of the original dipole. We call a shell *complete* if all its elements lie inside the volume of the scatterer $V$. A shell is called a *distant* shell if all its elements satisfy $R_{ij} > 1$, i.e. if its order $l > K_{\max} = \left[1/d\right]$. Let $K(i)$ be the order of the first incomplete shell, which is an indicator of how close dipole $i$ is to the surface. We demand $K(i) \le K_{\max}$ to separate cases (1) and (2) described above. All $j$ that fall in the third case satisfy $R_{ij} < 2$ (the exact value of this constant – slightly larger than $\sqrt{3}$ – depends on $d$). The number of dipoles in a shell $S_l$ (which can be incomplete) – $n_s(l)$ – can be estimated as

$$n_s(l) \le (2l+1)^3 - (2l-1)^3 \le c_{21} l^2 . \tag{41}$$

The sum of the error over all dipoles that lie in complete shells is then

$$\sum_{l=1}^{K(i)-1} \sum_{j \in S_l(i)} \left( \int_{V_j} d^3 r' \overline{\mathbf{G}}(\mathbf{r}_i, \mathbf{r}') \mathbf{P}(\mathbf{r}') - d^3 \overline{\mathbf{G}}_{ij}^{(0)} \mathbf{P}_j \right), \tag{42}$$

Since each shell in Eq. (42) is complete it can be divided into pairs of dipoles that are symmetric over the center of the shell ($j$ and $-j$). For convenience we set $\mathbf{r}_i = \mathbf{0}$. The inner sum in Eq. (42) can then be rewritten as

$$\frac{1}{2} \sum_{j \in S_l(i)} \left( \int_{V_j} d^3 r' \overline{\mathbf{G}}(\mathbf{r}') \big( \mathbf{P}(\mathbf{r}') + \mathbf{P}(-\mathbf{r}') \big) - d^3 \overline{\mathbf{G}}_{ij}^{(0)} \big( \mathbf{P}_j + \mathbf{P}_{-j} \big) \right), \tag{43}$$

Further we introduce the auxiliary function

$$\mathbf{u}(\mathbf{r}') = \frac{1}{2} \big( \mathbf{P}(\mathbf{r}') + \mathbf{P}(-\mathbf{r}') \big) - \mathbf{P}(\mathbf{0}), \tag{44}$$

which satisfies the following inequalities (follows from Eq. (30) for $\mathbf{P}(\mathbf{r})$ and Taylor series)

$$\left|\mathbf{u}(\mathbf{r})\right| \le c_{22} r^2, \left|\partial_\mu \mathbf{u}(\mathbf{r})\right| \le c_{23} r, \left|\partial_\mu \partial_\nu \mathbf{u}(\mathbf{r})\right| \le c_{24} \text{ for } \forall \mu, \nu . \tag{45}$$

Then Eq. (43) is equivalent to

$$\sum_{j \in S_l(i)} \left( \int_{V_j} d^3 r' \overline{\mathbf{G}}(\mathbf{r}') \mathbf{u}(\mathbf{r}') - d^3 \overline{\mathbf{G}}_{ij}^{(0)} \mathbf{u}_j \right) + \sum_{j \in S_l(i)} \left( \int_{V_j} d^3 r' \overline{\mathbf{G}}(\mathbf{r}') - d^3 \overline{\mathbf{G}}_{ij}^{(0)} \right) \mathbf{P}_i , \tag{46}$$

where $\mathbf{u}_j = \mathbf{u}(\mathbf{r}_j)$. To estimate the first term we apply Eq. (32) to the whole function under the integral. Using Eqs. (31) and (45) one may obtain

$$\max_{\mu\nu, \mathbf{r}' \in V_j} \left| \partial_\mu \partial_\nu \left( \overline{\mathbf{G}}(\mathbf{r}') \mathbf{u}(\mathbf{r}') \right) \right| \le c_{25} R_{ij}^{-3} , \tag{47}$$

and hence

$$\sum_{j \in S_l(i)} \left( \int_{V_j} d^3 r' \overline{\mathbf{G}}(\mathbf{r}') \mathbf{u}(\mathbf{r}') - d^3 \overline{\mathbf{G}}_{ij}^{(0)} \mathbf{u}_j \right) \le \sum_{j \in S_l(i)} c_{26} d^5 R_{ij}^{-3} \le c_{27} d^2 l^{-1} , \tag{48}$$

where we have used Eq. (41) and $R_{ij} \ge ld$ for $j \in S_l(i)$.

It is straightforward to show that

$$\sum_{j \in S_l(i)} \int_{V_j} d^3 r' \overline{\mathbf{G}}(\mathbf{r}') = \frac{2}{3} \overline{\mathbf{I}} \sum_{j \in S_l(i)} \int_{V_j} d^3 r' g(r') , \tag{49}$$

$$\sum_{j \in S_l(i)} \overline{\mathbf{G}}_{ij}^{(0)} = \frac{2}{3} \overline{\mathbf{I}} \sum_{j \in S_l(i)} g(R_{ij}) . \tag{50}$$

The derivation is based upon Eq. (2) and the equivalence $\dfrac{\hat{R}\hat{R}}{R^2} \Leftrightarrow \dfrac{1}{3}\overline{\mathbf{I}}$ in all sums and integrals that satisfy cubical symmetry. Then second part of Eq. (46) is transformed to

$$\left| \sum_{j \in S_l(i)} \left( \int_{V_j} d^3 r' \overline{\mathbf{G}}(\mathbf{r}') - d^3 \overline{\mathbf{G}}_{ij}^{(0)} \right) \mathbf{P}_i \right| \le c_{28} \sum_{j \in S_l(i)} \left| \int_{V_j} d^3 r' g(r') - d^3 g(R_{ij}) \right| \le c_{29} d^4 l + c_{30} d^2 l^{-3} , \tag{51}$$

where we apply Eq. (33) to derive the second inequality and use the identity $\nabla^2 g(r) = -g(r)$ and the following inequalities

$$\left| g(R) \right| \le c_{31} R^{-1}, \left| \partial_\mu \partial_\nu \partial_\rho \partial_\tau g(R) \right| \le c_{32} R^{-5} \text{ for } \forall \mu, \nu, \rho, \tau . \tag{52}$$

Substituting Eqs. (48) and (51) into Eq. (42) one can obtain

$$\sum_{l=1}^{K(i)-1} \sum_{j\in S_l(i)} \left( \int_{V_j} d^3r' \overline{\mathbf{G}}(\mathbf{r}_i,\mathbf{r}')\mathbf{P}(\mathbf{r}') - d^3 \overline{\mathbf{G}}_{ij}^{(0)} \mathbf{P}_j \right) \le \left(c_{33} + c_{34} \ln K(i)\right) d^2 , \tag{53}$$

using the fact that $K(i)d \le 1$.

We now consider the second part of the sum in Eq. (34) (where $R_{ij} > 1$). We first apply Eq. (32), then use Eq. (30) for $\mathbf{P}(\mathbf{r})$ and $\overline{\mathbf{G}}(\mathbf{r})$, and finally invoke Eq. (29):

$$\sum_{j,R_{ij}>1} \left( \int_{V_j} d^3r' \overline{\mathbf{G}}(\mathbf{r}_i,\mathbf{r}')\mathbf{P}(\mathbf{r}') - d^3 \overline{\mathbf{G}}_{ij}^{(0)} \mathbf{P}_j \right) \le \sum_{j,R_{ij}>1} c_{35} d^5 \le N c_{35} d^5 \le c_{36} d^2 . \tag{54}$$

To analyze the third part of the sum in Eq. (34) we again sum over shells, however since they are incomplete we cannot use symmetry considerations. We apply Eq. (33) to the whole function under the integral and proceed analogous to the derivation of Eq. (51). Using the identity

$$\nabla^2 \overline{\mathbf{G}}(\mathbf{r}) = -\overline{\mathbf{G}}(\mathbf{r}) , \tag{55}$$

(since we have assumed $k = 1$) we obtain

$$\left| \nabla^2 \left( \overline{\mathbf{G}}(\mathbf{r})\mathbf{P}(\mathbf{r}) \right)_{\mathbf{r}=\mathbf{R}_{ij}} \right| \le c_{37} R_{ij}^{-4} , \tag{56}$$

$$\max_{\mu\nu\rho\tau,\mathbf{r}'\in V_j} \left| \partial_\mu \partial_\nu \partial_\rho \partial_\tau \left( \overline{\mathbf{G}}(\mathbf{r}')\mathbf{P}(\mathbf{r}') \right) \right| \le c_{38} R_{ij}^{-7} , \tag{57}$$

which leads to

$$\sum_{j\in S_l(i)} \left( \int_{V_j} d^3r' \overline{\mathbf{G}}(\mathbf{r}_i,\mathbf{r}')\mathbf{P}(\mathbf{r}') - d^3 \overline{\mathbf{G}}_{ij}^{(0)} \mathbf{P}_j \right) \le c_{39} d l^{-2} + c_{40} l^{-5} , \tag{58}$$

and then analogous to Eq. (53):

$$\sum_{l=K(i)}^{K_{\max}} \sum_{j\in S_l(i)} \left( \int_{V_j} d^3r' \overline{\mathbf{G}}(\mathbf{r}_i,\mathbf{r}')\mathbf{P}(\mathbf{r}') - d^3 \overline{\mathbf{G}}_{ij}^{(0)} \mathbf{P}_j \right) \le c_{41} d K^{-1}(i) + c_{42} K^{-4}(i) . \tag{59}$$

Collecting Eqs. (40), (53), (54), (59) we finally obtain

$$\left|\mathbf{h}_i^d\right| \le c_{41} d K^{-1}(i) + c_{42} K^{-4}(i) + \left(c_{43} + c_{44} \ln K(i)\right) d^2 . \tag{60}$$

Then

$$\left\|\mathbf{h}^d\right\|_1 = \sum_{i=1}^{N} \left|\mathbf{h}_i^d\right| \le \left(c_{43} + c_{44} \ln K_{\max}\right) N d^2 + \sum_{K=1}^{K_{\max}} n(K) \left(c_{41} d K^{-1} + c_{42} K^{-4}\right), \tag{61}$$

where $n(K)$ is the number of dipoles whose order of the first incomplete shell is equal to $K$. It is clear that

$$n(K) \le n(1) \le \gamma_{12} N d , \tag{62}$$

where $\gamma_{12}$ is surface to volume ratio of the scatterer. Finally we obtain

$$\left\|\mathbf{h}^d\right\|_1 \le N\left[\left(c_{43} - c_{45} \ln d\right) d^2 + c_{46} d\right]. \tag{63}$$

The last term in Eq. (63) is mostly determined by dipoles that lie on the surface (or few dipoles deep) because it comes from the $K^{-4}$ term in Eq. (61) (which rapidly decreases when moving from surface). We define *surface errors* as those associated with the linear term in Eq. (63). Our numerical simulation (see Section 3.B) show that this term is small compared to other terms for “typical” values of $d$ , however it is always significant for small enough values of $d$.

From Eq. (18) we directly obtain

$$\left\|\delta\mathbf{E}^d\right\|_1 \le \left\|\left(\overline{\mathbf{A}}^d\right)^{-1}\right\|_1 \left\|\mathbf{h}^d\right\|_1 . \tag{64}$$

We assume that a bounded solution of Eq. (7) uniquely exists for any $\tilde{\mathbf{E}}^{inc} \in H_2$, moreover we assume that if $\left\|\tilde{\mathbf{E}}^{inc}\right\|_1 = 1$ then $\left\|\tilde{\mathbf{E}}\right\|_1 \le \gamma_{13}$. These assumptions are equivalent to the fact that $\left\|\tilde{\mathbf{A}}^{-1}\right\|_1$ exists and is finite (the operator $\tilde{\mathbf{A}}^{-1}$ is bounded). Because $\overline{\mathbf{A}}^d$ is a discretization of $\tilde{\mathbf{A}}$ one would expect that

$$\lim_{d\to 0}\left\|\left(\overline{\mathbf{A}}^d\right)^{-1}\right\|_1 = \left\|\tilde{\mathbf{A}}^{-1}\right\|_1 = \gamma_{13} . \tag{65}$$

Although Eq. (65) seems intuitively correct, its rigorous prove, even if feasible, lies outside the scope of this paper. For an intuitive understanding one may consult the paper by Rahola,[25] where he studied the spectrum of the discretized operator (for scattering by a sphere) and showed that it *does* converge to the spectrum of the integral operator with decreasing $d$. It should however be noted, that convergence of the spectrum only implies the convergence of the spectral ($L^2$) norm of the operator and not necessarily the convergence of the $L^1$-norm. Therefore Eq. (65) should be considered as an assumption. It implies that there exists a $d_0$ such that

$$\text{for } d < d_0 \quad \left\| \left( \overline{\mathbf{A}}^d \right)^{-1} \right\|_1 \le c_{47} , \tag{66}$$

where $c_{47}$ is an arbitrary constant larger then $\gamma_{13}$ (although $d_0$ depends on its choice). For example $c_{47} = 2\gamma_{13}$ should lead to a rather large $d_0$ (a rigorous estimate of $d_0$ does not seem feasible). Therefore $\left\| \delta \mathbf{E}^d \right\|_1$ satisfies the same constrain as $\left\| \mathbf{h}^d \right\|_1$ (Eq. (63)) but with constants $c_{48}$-$c_{50}$.

Next we estimate the errors in the measured quantities and start with the discretization error (first part in Eq. (28)). Examining Eqs. (20) and (24) one can see that Eq. (32) may be directly applied leading to

$$\left| \tilde{\phi}\left( \tilde{\mathbf{E}} \right) - \phi^d\left( \mathbf{E}^{0,d} \right) \right| \le \sum_i c_{51} d^5 \le c_{52} d^2 . \tag{67}$$

The second part in Eq. (28) is estimated as

$$\left| \phi^d\left( \mathbf{E}^{0,d} \right) - \phi^d\left( \mathbf{E}^d \right) \right| \le \sum_i c_{53} d^3 \left| \delta \mathbf{E}_i^d \right| \le c_{53} d^3 \left\| \delta \mathbf{E}^d \right\|_1 \le \left( c_{54} - c_{55} \ln d \right) d^2 + c_{56} d , \tag{68}$$

where we used Eq. (29). The estimation of the error for $C_{abs}$ additionally uses the fact $\left| \delta \mathbf{E}_i^d \right|^2 \le c_{57} \left| \delta \mathbf{E}_i^d \right| \quad \left( c_{57} = \max_{d<2,\, i} \left| \delta \mathbf{E}_i^d \right| \right)$.

By combining Eqs. (67) and (68) we obtain the final result of this section:

$$\left|\delta\phi^{d}\right| \leq \left(c_{58} - c_{55}\ln d\right)d^{2} + c_{56}d\ . \tag{69}$$

It is important to remember that the derivation was performed for constant $x$, $m$, shape, and incident field. There are 13 basic constants ($\gamma_1$-$\gamma_{13}$). $\gamma_1$ (Eq. (29)) characterizes the total volume of the scatterer, hence it depends only on $x$. $\gamma_7$-$\gamma_{11}$ (Eq. (30) for $\chi(\mathbf{r})$) can be easily obtained given the function $\chi(\mathbf{r})$, moreover it is completely trivial in the common case of homogenous scatterers. $\gamma_{12}$ (surface to volume ratio, Eq. (62)) depends on the shape of the scatterer and is inversely proportional to $x$. It is not feasible (except for certain simple shapes) to obtain the values of constants $\gamma_2$-$\gamma_6$ (Eq. (30)), since it requires an exact solution for the internal fields. These constants definitely depend upon all the parameters of the scattering problem. Moreover, these dependencies can be rapidly varying, especially near the resonance regions. The same is true for $\gamma_{13}$ ($L^1$-norm of the inverse of the integral operator, Eq. (65)). Finally, there is the important constant $d_0$ that also depends on all the parameters, however one may expect it to be large enough (e.g. $d_0 \geq 2$) for most of the problems – then its variation can be neglected.

Before proceeding we introduce the discretization parameter $y = |m|kd$. We employ the commonly used formula as proposed by Draine,[8] however the exact dependence on $m$ is not important because all the conclusions are still valid for constant $m$. Replacing $d$ by $y$ does not significantly change the dependence of the constants in Eq. (69) since they all already depend on $m$ through the basic constants $\gamma_2$-$\gamma_{11}$, $\gamma_{13}$. This leads to

$$\left|\delta\phi^{y}\right| \leq \left(c_{59} - c_{60}\ln y\right)y^{2} + c_{61}y\ . \tag{70}$$

It is not feasible to make any rigorous conclusions about the variation of the constants in Eq. (70) with varying parameters because all these constants depend on $\gamma_2$-$\gamma_6$, $\gamma_{13}$ that in turn depend in a complex way upon the parameters of the scattering problem. However we can make one conclusion about the general trend of this dependency.

Following the derivation of the Eq. (70) one can observe that $c_{61}$ is proportional to $\gamma_{12}$, while $c_{59}$ and $c_{60}$ do not directly depend on it (at least part of the contributions to them are independent of $\gamma_{12}$). Therefore the general trend will be a decrease of the ratio $c_{61}/c_{59}$ with increasing $x$ (when all other parameters are fixed). This is a mathematical justification of the intuitively evident fact that surface errors are less significant for larger particles.

In the analysis of the results of the numerical simulations (Section 3.B) we will neglect the variation of the logarithm. Eq. (70) then states that error is bounded by a quadratic function of $y$ (for $d \le d_0$). However, keep in mind that our derivation does not lead to an optimal error estimation, i.e. it overestimates the error and can be improved. For example, the constants $\gamma_2$-$\gamma_6$ are usually largest inside a small volume fraction of the scatterer (near the surface or some internal resonance regions), while in the rest of the scatterer the internal electric field and its derivatives are bounded by significantly smaller constants. However the order of the error is estimated correctly, as we will see in the numerical simulations.

It is important to note that Eq. (70) does not imply that $\delta\phi^y$ (which is a signed value) actually depends on $y$ as a quadratic function, but we will see later that it is the case for small enough $y$ (Section 3.B, see detailed discussion in Paper 2). Moreover, the coefficients of linear and quadratic terms for $\delta\phi^y$ may have different signs, which may lead to zero error for non-zero $y$ (however, this $y$, if it exists, is unfortunately different for each measured quantity).

### *E.Shape errors*

In this section we extend the error analysis as presented in Section 2.D to shapes that cannot be described exactly by a set of cubical subvolumes.

We perform the discretization the same way as in Section 2.B but some of the $V_i$ are not cubical (for $i \in \partial V$, which denotes that dipole $i$ lies on the boundary of the volume $V$). We set $\mathbf{r}_i$ to be still in the center of the cube (circumscribing $V_i$) not to break the regularity of the lattice. The standard PP prescription uses equal volumes ($V_i = d^3$) in Eqs. (10), (14), (25), and (26), i.e. the discretization changes the shape of the particle a little bit. We will estimate the errors introduced by these boundary dipoles. These errors should then be added to those obtained in Section 2.D. We start by estimating $\left\|\mathbf{h}^d\right\|_1$. First we consider $\mathbf{h}_i^d$ for $i \notin \partial V$

$$\mathbf{h}_i^d = \sum_{j\in\partial V}\left(\int_{V_j} d^3 r' \overline{\mathbf{G}}(\mathbf{r}_i,\mathbf{r}')\mathbf{P}(\mathbf{r}') - d^3 \overline{\mathbf{G}}_{ij}^{(0)}\mathbf{P}_j\right), \tag{71}$$

which is just a reduction of Eq. (34). For $i \in \partial V$ $\mathbf{h}_i^d$ is the same plus the error in the self-term

$$\mathbf{h}_i^d = \sum_{\substack{j\in\partial V\\ j\neq i}}\left(\int_{V_j} d^3 r' \overline{\mathbf{G}}(\mathbf{r}_i,\mathbf{r}')\mathbf{P}(\mathbf{r}') - d^3 \overline{\mathbf{G}}_{ij}^{(0)}\mathbf{P}_j\right) + \mathbf{M}(V_i,\mathbf{r}_i) - \left(\overline{\mathbf{L}}(\partial V_i,\mathbf{r}_i) - \frac{4\pi}{3}\overline{\mathbf{I}}\right)\chi_i \mathbf{E}_i . \tag{72}$$

Let us define

$$\mathbf{h}_{ij}^{sh} = \int_{V_j} d^3 r' \overline{\mathbf{G}}(\mathbf{r}_i,\mathbf{r}')\mathbf{P}(\mathbf{r}') - d^3 \overline{\mathbf{G}}_{ij}^{(0)}\mathbf{P}_j , \tag{73}$$

$$\mathbf{h}_{ii}^{sh} = \mathbf{M}(V_i,\mathbf{r}_i) - \left(\overline{\mathbf{L}}(\partial V_i,\mathbf{r}_i) - \frac{4\pi}{3}\overline{\mathbf{I}}\right)\chi_i \mathbf{E}_i . \tag{74}$$

We estimate each of the terms in Eq. (73) separately (since there is actually no significant cancellation, and the error is of the same order of magnitude as the values themselves) using Eq. (30) for $\mathbf{P}(\mathbf{r})$ and $\overline{\mathbf{G}}(\mathbf{r})$ and Eq. (31). This leads to

$$\left|\mathbf{h}_{ij}^{sh}\right| \le \begin{cases} c_{62} d^3 R_{ij}^{-3},\ R_{ij} < 2 \\ c_{63} d^3,\ R_{ij} > 1 \end{cases} . \tag{75}$$

To estimate $\mathbf{h}_{ii}^{sh}$ we assume that the surface of the scatterer is a plane on the scale of the size of the dipole. A finite radius of curvature only changes the constants in the following expressions. We will prove that

$$\left|\mathbf{h}_{ii}^{sh}\right| \le c_{64} , \tag{76}$$

therefore we do not need to consider the third term in Eq. (74) (coming from the unity tensor) at all, since it is bounded by a constant.

$$\mathbf{M}(V_i,\mathbf{r}_i)=\int_{V_i} d^3r'\left(\overline{\mathbf{G}}(\mathbf{r}_i,\mathbf{r}')-\overline{\mathbf{G}}^s(\mathbf{r}_i,\mathbf{r}')\right)\mathbf{P}(\mathbf{r}')+\int_{V_i} d^3r'\overline{\mathbf{G}}^s(\mathbf{r}_i,\mathbf{r}')\left(\mathbf{P}(\mathbf{r}')-\mathbf{P}(\mathbf{r}_i)\right). \tag{77}$$

The function in the first integral is always bounded by $c_{65}\left|\mathbf{r}'-\mathbf{r}_i\right|^{-2}$. If $\mathbf{r}_i\in V_i$ the same is true for the second integral and hence

$$\left|\mathbf{M}(V_i,\mathbf{r}_i)\right|\leq c_{66}d\,. \tag{78}$$

If $\mathbf{r}_i\notin V_i$ we introduce an auxiliary point $\mathbf{r}''$ that is symmetric to $\mathbf{r}_i$ over the particle surface and apply the identity

$$\mathbf{P}(\mathbf{r}')-\mathbf{P}(\mathbf{r}_i)=\left(\mathbf{P}(\mathbf{r}')-\mathbf{P}(\mathbf{r}'')\right)+\left(\mathbf{P}(\mathbf{r}'')-\mathbf{P}(\mathbf{r}_i)\right) \tag{79}$$

to the second integral in Eq. (77). Using Taylor expansion of $\mathbf{P}$ near $\mathbf{r}''$ and the fact that $\left|\mathbf{r}'-\mathbf{r}''\right|\leq\left|\mathbf{r}'-\mathbf{r}_i\right|$ for $\mathbf{r}'\in V_i$ one can show that

$$\left|\mathbf{M}(V_i,\mathbf{r}_i)\right|\leq c_{67}d+c_{68}\left|\int_{V_i} d^3r'\overline{\mathbf{G}}^s(\mathbf{r}_i,\mathbf{r}')\right|, \tag{80}$$

where the remaining integral can be proven to be equal to $-\overline{\mathbf{L}}(\partial V_i,\mathbf{r}_i)$. The last prove left (see Eqs. (74) and (80)) is to demonstrate that $\overline{\mathbf{L}}(\partial V_i,\mathbf{r}_i)$ is bounded by a constant. The only potential problem may come from the subsurface of $\partial V_i$ that is part of the particle surface (because it may be close to $\mathbf{r}_i$). This subsurface is assumed planar. We will calculate the integral in Eq. (6) over the infinite plane $\mathbf{r}'-\mathbf{r}_i=\boldsymbol{\rho}+\mathbf{r}$ such that $\boldsymbol{\rho}\cdot\mathbf{r}=0$. Then $\mathbf{n}'=\pm\boldsymbol{\rho}/\rho$ and

$$\overline{\mathbf{L}}(\text{inf.plane},\mathbf{r}_i)=\mp\oint_{\partial V_0} d^2r\frac{\hat{\rho}\hat{\rho}}{\rho\left(\rho^2+r^2\right)^{3/2}}=\mp 2\pi\frac{\hat{\rho}\hat{\rho}}{\rho^2}, \tag{81}$$

which is bounded. The rest of the integral (over the part of the cube surface) is bounded by a constant, which is a manifestation of a more general fact that (by its definition) $\overline{\mathbf{L}}(\partial V_i, \mathbf{r}_i)$ does not depend on the size but only on the shape of the volume. Finally we have

$$\left|\overline{\mathbf{L}}(\partial V_i, \mathbf{r}_i)\right| \leq c_{69}, \tag{82}$$

which together with Eqs. (74), (78), and (80) prove Eq. (76).

Using Eqs. (75) and (76) we obtain

$$\left\|\mathbf{h}^d\right\|_1 \leq \sum_i \sum_{j\in\partial V}\left|\mathbf{h}_{ij}^{sh}\right| + \sum_{i\in\partial V}\left|\mathbf{h}_{ii}^{sh}\right| \leq \sum_{j\in\partial V}\left(\sum_{l=1}^{K_{\max}} c_{62} n_s(l) l^{-3} + c_{70}\right) \leq Nd(c_{71} - c_{72}\ln d), \tag{83}$$

where we have changed the order of the summation in the double sum and split the summation over cubical shells for $l \leq K_{\max}$ and $l > K_{\max}$. Then we have grouped everything into one sum over boundary dipoles. Eqs. (41) and (62) were used in the last inequality. Combining Eqs. (63) and (83) one can obtain the total estimate of the $\left\|\mathbf{h}^d\right\|_1$ for any scatterer:

$$\left\|\mathbf{h}^d\right\|_1 \leq N\left[(c_{43} - c_{45}\ln d)d^2 + (c_{73} - c_{72}\ln d)d\right]. \tag{84}$$

Using Eq. (66) we immediately obtain the same estimate for $\left\|\delta\mathbf{E}^d\right\|_1$.

The derivation of the errors in the measured quantities is slightly modified compared to Section 2.D, by the presence of the shape errors. Eqs. (67) and (68) are changed to

$$\left|\tilde{\phi}\left(\tilde{\mathbf{E}}\right) - \phi^d\left(\mathbf{E}^{0,d}\right)\right| \leq \sum_i c_{51} d^5 + \sum_{i\in\partial V} c_{74} d^3 \leq c_{52} d^2 + c_{75} d, \tag{85}$$

$$\left|\phi^d\left(\mathbf{E}^{0,d}\right) - \phi^d\left(\mathbf{E}^d\right)\right| \leq c_{53} d^3 \left\|\delta\mathbf{E}^d\right\|_1 \leq (c_{54} - c_{55}\ln d)d^2 + (c_{76} - c_{77}\ln d)d. \tag{86}$$

The second term in Eq. (85) comes from surface dipoles for which errors are the same order as the values themselves. Finally the generalization of Eq. (70) is

$$\left|\delta\phi^y\right| \leq (c_{59} - c_{60}\ln y)y^2 + (c_{78} - c_{79}\ln y)y. \tag{87}$$

The shape errors "reinforce" the surface errors (the linear term of discretization error), and although they both generally decrease with increasing size parameter $x$ one may expect the linear term in Eq. (87) to be significant up to higher values of $y$ than in Eq. (70).

All the derivations in this section can in principle be extended to interfaces inside the particle, i.e. when a surface, which cannot be described exactly as a surface of a set of cubes, separates two regions where $\chi(\mathbf{r})$ varies smoothly. Two parts of the cubical dipole on the interface should be considered separately the same way as it was done above. This will however not change the main conclusion of this section – Eq.(87) – but only the constants.

### F. Different DDA formulations

In this section we discuss how different DDA formulations modify the error estimates derived in Sections 2.D and 2.E.

Most of the improvements of PP proposed in the literature are concerned with the self-term – $\mathbf{M}(V_i,\mathbf{r}_i)$. They are the Radiative Reaction correction (RR),[8] the Digitized Green's Function (DGF),[23] the formulation by Lakhtakia (LAK),[26,27] the $a_1$-term method,[28,29] the Lattice Dispersion Relation (LDR),[30] the formulation by Peltoniemi (PEL),[31] and the Corrected LDR (CLDR).[32] All of them provide an expression for $\mathbf{M}(V_i,\mathbf{r}_i)$ that is of order $d^2$ (except for RR that is of order $d^3$). For instance LDR is equivalent to

$$\mathbf{M}(V_i,\mathbf{r}_i) = \left[\left(b_1 + b_2 m^2 + b_3 m^2 S\right)d^2 + (2/3)\mathrm{i}d^3\right]\mathbf{P}_i, \tag{88}$$

(remember that we assumed $k=1$) where $b_1$, $b_2$, $b_3$ are numerical constants and $S$ is a constant that depends only on the propagation and polarization vectors of the incident field. However, none of these formulations can exactly evaluate the integral in Eq. (39), because the variation of the electric field is not known beforehand (PEL solves this problem, but only for a spherical dipole). Therefore they (hopefully) decrease the constant in Eq. (40), thus decreasing the overall error in the measured quantities. However, these formulations are not expected to change the order of the error from $d^2$ to some higher order.

We do not analyze the improvements by Rahmani, Chaumet, and Bryant (RCB)[33,34] and Surface Corrected LDR (SCLDR),[17] as they are limited to certain particle shapes.

There exist two improvements of the interaction term in PP: Filtered Coupled Dipoles (FCD)[12] and Integration of Green's Tensor (IT).[35] A rigorous analysis of FCD errors is beyond the scope of this paper, but it seems that FCD is not designed to reduce the linear term in Eq. (63) that comes from the incomplete (non-symmetric) shells. This is because FCD employs sampling theory to improve the accuracy of the overall discretization for regular cubical grids. FCD does not improve the accuracy of a single $\overline{\mathbf{G}}_{ij}$ calculation (approximation of an integral over one subvolume).

IT, which numerically evaluates the integral in Eq. (12), has a more pronounced effect on the error estimate. Consider dipole $j$ from $l$-th shell (incomplete) of dipole $i$, then

$$\begin{aligned}
&\int_{V_j} d^3r'\overline{\mathbf{G}}(\mathbf{r}_i,\mathbf{r}')\mathbf{P}(\mathbf{r}') - d^3\overline{\mathbf{G}}_{ij}\mathbf{P}_j = \int_{V_j} d^3r'\overline{\mathbf{G}}(\mathbf{r}_i,\mathbf{r}')\left(\mathbf{P}(\mathbf{r}') - \mathbf{P}_j\right) \le \\
&c_{80}\int_{V_j} d^3r' r'^2 \max_{\mu,r'\in V_j}\left|\partial\overline{\mathbf{G}}(\mathbf{r}_i,\mathbf{r}')\right| + c_{81}\int_{V_j} d^3r'\left|\overline{\mathbf{G}}(\mathbf{r}_i,\mathbf{r}')\right| r'^2 \le c_{82}dl^{-4}.
\end{aligned} \tag{89}$$

Here we have used Eq. (36) and Taylor expansion of Green's tensor up to the first order. Eq. (89) states that the second term in Eq. (58) is completely eliminated and so is the linear term in Eqs. (69) and (70) (surface errors). Therefore convergence of DDA with IT for cubically shaped scatterers is expected to be purely quadratic (neglecting the logarithm). However, for non cubically shaped scatterers the linear term reappears, due to the shape errors. Both IT and FCD also modify the self-term, however the effect is basically the same as for the other formulations.

Several papers aimed to reduce shape errors.[10,11,36] The first one – Generalized Semi-Analytical (GSA) method[10] – modifies the whole DDA scheme, while the other two propose averaging of the susceptibility over the boundary dipoles. We will analyze here Weighted Discretization (WD) by Piller[11] as probably the most advanced method to reduce shape errors available today.

WD modifies the susceptibility and self-term of the boundary subvolume. We slightly modify the definition of the boundary subvolume used in Sections 2.B and 2.E to automatically take into account interfaces inside the scatterer. We define $V_i$ to be always cubical, but with a possible interface inside. The particle surface, crossing the subvolume $V_i$, is assumed planar and divides the subvolume into two parts: the principal volume $V_i^p$ (containing the center) and the secondary volume $V_i^s$ with susceptibilities $\chi_i^p \equiv \chi_i$, $\chi_i^s$ and electric fields $\mathbf{E}_i^p \equiv \mathbf{E}_i$, $\mathbf{E}_i^s$ respectively. The electric fields are considered constant inside each part and related to each other via the boundary condition tensor $\overline{\mathbf{T}}_i$:

$$\mathbf{E}_i^s = \overline{\mathbf{T}}_i \mathbf{E}_i . \qquad (90)$$

In WD the susceptibility of the boundary subvolume is replaced by an effective one, defined as

$$\overline{\chi}_i^e = \left( V_i^p \chi_i^p \overline{\mathbf{I}} + V_i^s \chi_i^s \overline{\mathbf{T}}_i \right) / d^3 , \qquad (91)$$

which gives the correct total polarization of the cubical dipole. The effective self-term is directly evaluated starting from Eq. (4), considering $\chi$ and $\mathbf{E}$ constant inside each part,

$$\overline{\mathbf{M}}_i^e \overline{\boldsymbol{\chi}}_i^e \mathbf{E}_i = \left( \int_{V_i^p} \mathrm{d}^3 r' \left( \overline{\mathbf{G}}(\mathbf{r}_i, \mathbf{r}') - \overline{\mathbf{G}}^s(\mathbf{r}_i, \mathbf{r}') \right) \chi_i^p + \int_{V_i^s} \mathrm{d}^3 r' \left( \overline{\mathbf{G}}(\mathbf{r}_i, \mathbf{r}') \chi_i^s \overline{\mathbf{T}}_i - \overline{\mathbf{G}}^s(\mathbf{r}_i, \mathbf{r}') \chi_i^p \right) \right) \mathbf{E}_i . \qquad (92)$$

Piller evaluated the integrals in Eq. (92) numerically.[11]

To take a smooth variation of the electric field and susceptibility into account we define $\chi_i^s = \chi(\mathbf{r}'')$ ($\mathbf{r}''$ is defined in Section 2.E) and $\overline{\mathbf{T}}_i$ is calculated at the surface between $\mathbf{r}_i$ and $\mathbf{r}''$. $\mathbf{P}_i^p \equiv \mathbf{P}_i$ and $\mathbf{P}_i^s = \chi_i^s \mathbf{E}_i^s = \chi_i^s \overline{\mathbf{T}}_i \mathbf{E}_i$. Then

$$\left|\mathbf{P}(\mathbf{r}'') - \mathbf{P}_i^s\right| \leq c_{83} \min_{\mathbf{r}\in V_i^s} \left|\mathbf{r} - \mathbf{r}_i\right|, \tag{93}$$

where we have assumed that Eq. (30) for $\chi(\mathbf{r})$ and $\mathbf{E}(\mathbf{r})$ is also valid in $V_i^s$.

We start estimating errors of WD with $\mathbf{h}_{ij}^{sh}$ (cf. Eq. (73))

$$\mathbf{h}_{ij}^{sh} = \int_{V_j^p} d^3r' \left(\overline{\mathbf{G}}(\mathbf{r}_i, \mathbf{r}')\mathbf{P}(\mathbf{r}') - \overline{\mathbf{G}}_{ij}^{(0)}\mathbf{P}_j^p\right) + \int_{V_j^s} d^3r' \left(\overline{\mathbf{G}}(\mathbf{r}_i, \mathbf{r}')\mathbf{P}(\mathbf{r}') - \overline{\mathbf{G}}_{ij}^{(0)}\mathbf{P}_j^s\right), \tag{94}$$

Using Taylor expansions of $\mathbf{P}(\mathbf{r}')$ near $\mathbf{r}_i$ and $\mathbf{r}''$ in $V_i^p$ and $V_i^s$ correspondingly and Eq. (93) one may obtain that the main contribution comes from the derivative of Green's tensor, leading to (cf. Eq. (75))

$$\left|\mathbf{h}_{ij}^{sh}\right| \leq \begin{cases} c_{84} d^4 R_{ij}^{-4}, \; R_{ij} < 2; \\ c_{85} d^4, \; R_{ij} > 1. \end{cases} \tag{95}$$

$\mathbf{h}_{ii}^{sh}$ is the following (cf. Eq. (74))

$$\begin{aligned} \mathbf{h}_{ii}^{sh} &= \left(\mathbf{M}(V_i, \mathbf{r}_i) - \overline{\mathbf{L}}(\partial V_i, \mathbf{r}_i)\mathbf{P}_i^p\right) - \\ &- \left( \int_{V_i^p} \mathrm{d}^3r' \left(\overline{\mathbf{G}}(\mathbf{r}_i, \mathbf{r}') - \overline{\mathbf{G}}^s(\mathbf{r}_i, \mathbf{r}')\right)\mathbf{P}_i^p + \int_{V_i^s} \mathrm{d}^3r' \left(\overline{\mathbf{G}}(\mathbf{r}_i, \mathbf{r}')\mathbf{P}_i^s - \overline{\mathbf{G}}^s(\mathbf{r}_i, \mathbf{r}')\mathbf{P}_i^p\right) - \overline{\mathbf{L}}(\partial V_i, \mathbf{r}_i)\mathbf{P}_i^p \right) \\ &= \int_{V_i^p} \mathrm{d}^3r' \overline{\mathbf{G}}(\mathbf{r}_i, \mathbf{r}')\left(\mathbf{P}(\mathbf{r}') - \mathbf{P}_i^p\right) + \int_{V_i^s} \mathrm{d}^3r' \overline{\mathbf{G}}(\mathbf{r}_i, \mathbf{r}')\left(\mathbf{P}(\mathbf{r}') - \mathbf{P}_i^s\right). \end{aligned} \tag{96}$$

The first two integrals can be easily shown to be $\leq c_{85} d$ (cf. Eq. (77)), thus

$$\left|\mathbf{h}_{ii}^{sh}\right| \leq c_{86} d \,. \tag{97}$$

Proceeding analogous to the derivation of Eq. (83) one can obtain

$$\left\|\mathbf{h}^d\right\|_1 \leq \sum_{j\in\partial V} \left( \sum_{l=1}^{K_{\max}} c_{84} n_s(l) l^{-4} + c_{88} d \right) \leq c_{89} N d \,. \tag{98}$$

It can be shown that for the scattering amplitude (Eq. (25)) the error estimate given by Eq. (85) can be improved, since WD correctly evaluates the zeroth order of value for the boundary dipoles, leading to

$$\left|\tilde{\phi}\left(\tilde{\mathbf{E}}\right)-\phi^{d}\left(\mathbf{E}^{0,d}\right)\right| \leq \sum_{i} c_{51} d^{5}+\sum_{i \in \partial V} c_{90} d^{4} \leq c_{91} d^{2}. \tag{99}$$

In his original paper[11] Piller did not specify the expression that should be used for $C_{abs}$. Direct application of the susceptibility provided by WD into Eq. (26) does not reduce the order of error when compared with the exact Eq. (24) (except when $\chi_i^s = 0$), since they are not linear functions of the electric field. However, if we consider separately $V_i^p$ and $V_i^s$ (which is equivalent to replacing $V_i \operatorname{Im}\left(\left(\overline{\chi}_i^e \mathbf{E}_i\right) \cdot \mathbf{E}_i^*\right)$ by $V_i^p \operatorname{Im}(\chi_i^p)\left|\mathbf{E}_i\right|^2 + V_i^s \operatorname{Im}(\chi_i^s)\left|\overline{\mathbf{T}}_i \mathbf{E}_i\right|^2$ ) the same estimate as in Eq. (99) can be derived for $C_{abs}$.

Using Eqs. (98), (99), and the first part of Eq. (86) one can derive the final error estimate for WD:

$$\left|\delta\phi^{y}\right| \leq \left(c_{92} - c_{93} \ln y\right) y^{2} + c_{94} y, \tag{100}$$

where the constant before the linear term, as compared to Eq. (87), does not contain a logarithm and is expected to be significantly smaller, because several factors contributing to it are eliminated in WD. Although WD has a potential for improving, it does not seem feasible to completely eliminate the linear term in the shape error. The accuracy of evaluation of the interaction term over the boundary dipole (cf. Eq. (94)) can be improved by integration of Green's tensor over $V_i^p$ and $V_i^s$ separately but that would ruin the block-Toeplitz structure of the interaction matrix and hinder the FFT-based algorithm for the solution of linear equations.[5] Since there is no comparable alternative to FFT nowadays, this method seems inapplicable.

Minor modifications of the expression for $C_{abs}$ are possible. Draine[8] proposed a modification of Eq. (26) that was widely used afterwards and which was further modified by Chaumet *et al.*[35] However, for many cases these expressions are equivalent and, even when they are not, the difference is of order $d^3$, which is neglected in our error analysis.

## 3. Numerical simulations

### A. Discrete Dipole Approximation

The basics of the DDA method were summarized by Draine and Flatau.[2] In this paper we use the LDR prescription for dipole polarizability,[30] which is most widely used nowadays, e.g. in the publicly available code DDSCAT 6.1.[4] We also employ dipole size correction[8] for non-cubically shaped scatterers to ensure that the cubical approximation of the scatterer has the correct volume; this is believed to diminish shape errors, especially for small scatterers.[2] We use a standard discretization scheme as described in Section 2.E, without any improvements for boundary dipoles. It is important to note that all the conclusions are valid for any DDA implementation, but with a few changes for specific improvements as discussed in Section 2.F.

Our code – Amsterdam DDA (ADDA) – is capable of running on a cluster of computers (parallelizing a *single* DDA computation), which allows us to use practically unlimited number of dipoles, since we are not limited by the memory of a single computer.[37,38] We used a relative error of residual $<10^{-8}$ as a stopping criterion for the iterative solution of the DDA linear system. Tests suggest that the relative error of the measured quantities due to the iterative solver is then $<10^{-7}$ (data not shown) and hence can be neglected (total relative errors in our simulations are $>10^{-6} \div 10^{-5}$ – see Section 3.B). More details about our code can be found in Paper 2. All DDA simulations were carried out on the Dutch national compute cluster LISA.[39]

### B. Results

We study five test cases: one cube with $kD = 8$, three spheres with $kD = 3, 10, 30$, and a particle obtained by a cubical discretization of the $kD = 10$ sphere using 16 dipoles per $D$ (total 2176 dipoles, $x$ equal to that of a sphere; see detailed description in Paper 2). By $D$ we denote the diameter of a sphere or the edge size of a cube. All scatterers are homogenous with $m = 1.5$. Although DDA errors significantly depend on $m$ (see e.g. [14]), we limit ourselves to one single value and study effects of size and shape of the scatterer.

The maximum number of dipoles per $D$ ($n_D$) was 256. The values of $n_D$ that we used are of the form $\{4,5,6,7\} \cdot 2^p$ ($p$ is an integer), except for the discretized sphere, where all $n_D$ are multiples of 16 (this is required to exactly describe the shape of the particle composed from a number of cubes). The minimum values for $n_D$ were 8 for the $kD = 3$ sphere, 16 for the cube, the $kD = 10$ sphere, and the discretized sphere, and 40 for the $kD = 30$ sphere.

All the computations use a direction of incidence parallel to one of the principal axes of the cubical dipoles. The scattering plane is parallel to one of the face of the cubical dipoles. In this paper we show results only for the extinction efficiency $Q_{ext}$ (for incident light polarized parallel to one of the principal axes of the cubical dipoles) and phase function $S_{11}(\theta)$ as the most commonly used in applications. However, the theory applies to any measured quantity. For instance, we have also confirmed it for other Mueller matrix elements (data not shown).

Exact results of $S_{11}(\theta)$ for all 5 test cases are shown in Fig. 2. For spheres this is the result of Mie theory (the relative accuracy of the code we used[24] is at least $< 10^{-6}$) and for the cube and discretized sphere an extrapolation over the 5 finest discretizations (the extrapolation technique is presented in Paper 2, together with all details of obtaining these results, including their estimated errors). We use such 'exact' results because analytical theory is unavailable for these shapes and because errors of the best discretization are larger than that of the extrapolation. Their use as references for computing real errors (difference between the computed and the exact value) of single DDA calculations is justified because all these real errors are significantly larger than the errors of the references themselves (see Paper 2; in general, real errors obtained this way have an uncertainty of reference error). Exact values of $Q_{ext}$ for all test cases are presented in Table 1.

In the following we show the results of DDA convergence. Fig. 3 and Fig. 4 present relative errors (absolute values) of $S_{11}$ at different angles θ and maximum error over all θ versus $y$ in log-log scale. In many cases the maximum errors are reached at exact backscattering direction, then these two sets of points overlap. Deep minima that happen at intermediate values of $y$ for some values of θ (and also sometimes for $Q_{ext}$ – Fig. 5) are due to the fact that the differences between simulated and reference values change sign near these values of $y$ (see Paper 2 for detailed description of this behavior). The solid lines are linear fits to all or some points of maximum error. The slopes of these lines are depicted in the figures. Fig. 5 shows relative errors of $Q_{ext}$ for all 5 studied cases in log-log scale. A linear fit through the 5 finest discretizations of the $kD = 3$ sphere is shown. More results of these numerical simulations are presented in Paper 2.

## 4. Discussion

Convergence of DDA for cubically shaped particles (Fig. 3) shows the following trends. All curves have linear and quadratic parts (the non-monotonic behavior of errors for some θ are also a manifestation of the fact that signed difference can be approximated by a sum of linear and quadratic terms that have different signs). The transition between these two regimes occurs at different $y$ (which indicates the relative importance of linear and quadratic coefficients). While for maximum errors that are close to those of the backscattering direction the linear term is significant for larger $y$, it is much smaller and not significant in the whole range of $y$ studied for side scattering ( $\theta = 90°$ ). Results of DDA convergence for spheres (Fig. 4) show a different behavior for different sizes. Errors for the small ( $kD = 3$ ) sphere converge purely linear (except for small deviation of errors of $S_{11}(90°)$ for large values of $y$). Similar results are obtained for the $kD = 10$ sphere, but with significant oscillations superimposed upon the general trend. Convergence for the large ($kD = 30$) sphere is quadratic or even faster in the range of $y$ studied, also with significant oscillations.

Comparing Fig. 3 and Fig. 4 (especially Fig. 3(b) and Fig. 4(b) showing results for almost the same particles) one can deduce the following differences in DDA convergence for cubically and non-cubically shaped scatterers. The linear term for cubically shaped scatterers is significantly smaller, resulting in smaller total errors, especially for small $y$. All these conclusions, together with the size dependence of the significance of the linear term in the total errors, are in perfect agreement with the theoretical predictions made in Sections 2.D and 2.E. Errors for non-cubically shaped particles exhibit quasi-random oscillations that are not present for cubically shaped particles. This can be explained by the sharp variations of shape errors with changing $y$ (discussed in details in Paper 2). Oscillations for the $kD = 3$ sphere (Fig. 4(a)) are very small (but still clearly present), which is due to the small size of the particle and hence featurelessness of its light scattering pattern – the surface structure is not that important and one may expect rather small shape errors. Results for $Q_{ext}$ (Fig. 5) fully support the conclusions. Errors of $Q_{ext}$ for the large sphere at small values of $y$ are unexpectedly smaller than for smaller spheres. This feature requires further study before making any firm conclusions, however there is definitely no similar tendency for $S_{11}(\theta)$ (cf. Fig. 4).

We have also studied a $kD = 8$ porous cube that was obtained by dividing a cube into 27 smaller cubes and then removing randomly 9 of them. All the conclusions are the same as those reported for the cube, but with slightly higher overall errors (data not shown).

In this paper have used a traditional DDA formulation[2] for numerical simulations. However, as we showed in Section 2.F several modern improvements of DDA (namely IT and WD) should significantly change its convergence behavior. IT should completely eliminate the linear term for cubically shaped scatterers, which should improve the accuracy especially for small $y$. WD should significantly decrease shape and hence total errors for non-cubically shaped particles, moreover it should significantly decrease the amplitude of quasi-random error oscillations because it takes into account the location of the interface inside the boundary dipoles. Numerical testing of DDA convergence using IT and WD is a subject of a future study.

## 5. Conclusion

To the best of our knowledge, we conducted for the first time a rigorous theoretical convergence analysis of DDA. In the range of DDA applicability ( $kd < 2$ ) errors are bounded by a sum of a linear and quadratic term in the discretization parameter $y$; the linear term is significantly smaller for cubically than for non-cubically shaped scatterers. Therefore for small $y$ errors for cubically shaped particles are much smaller than for non-cubically shaped. The relative importance of the linear term decreases with increasing size, hence convergence of DDA for large enough scatterers is quadratic in the common range of $y$. All these conclusions were verified by extensive numerical simulations.

Moreover, these simulations showed that errors are not only bounded by a quadratic function (as predicted in Section 2), but actually can be (with good accuracy) described by a quadratic function of $y$. This fact provides a basis for the extrapolation technique presented in Paper 2.

Our theory predicts that modern DDA improvements (namely IT and WD) should significantly change the convergence of DDA, however numerical testing of these predictions is left for future research.

## Acknowledgements

We thank Gorden Videen and Michiel Min for valuable comments on earlier version of this manuscript and anonymous reviewer for helpful suggestions. Our research is supported by the NATO Science for Peace program through grant SfP 977976.

## References

1. E. M. Purcell and C. R. Pennypacker, "Scattering and adsorption of light by nonspherical dielectric grains," Astrophys. J. **186**, 705-714 (1973).

2. B. T. Draine and P. J. Flatau, "Discrete-dipole approximation for scattering calculations," J. Opt. Soc. Am. A **11**, 1491-1499 (1994).

3. B. T. Draine, "The discrete dipole approximation for light scattering by irregular targets," in *Light Scattering by Nonspherical Particles, Theory, Measurements, and Applications,* M. I. Mishchenko, J. W. Hovenier, and L. D. Travis, eds., (Academic Press, 2000), pp. 131-145.

4. B. T. Draine and P. J. Flatau "User guide for the discrete dipole approximation code DDSCAT 6.1," http://xxx.arxiv.org/abs/astro-ph/0409262 (2004).

5. J. J. Goodman, B. T. Draine, and P. J. Flatau, "Application of fast-fourier-transform techniques to the discrete-dipole approximation," Opt. Lett. **16**, 1198-1200 (1991).

6. G. C. Hsiao and R. E. Kleinman, "Mathematical foundations for error estimation in numerical solutions of integral equations in electromagnetics," IEEE Trans. Ant. Propag. **45**, 316-328 (1997).

7. K. F. Warnick and W. C. Chew, "Error analysis of the Moment Method," IEEE Ant. Prop. Mag. **46**, 38-53 (2004).

8. B. T. Draine, "The discrete-dipole approximation and its application to interstellar graphite grains," Astrophys. J. **333**, 848-872 (1988).

9. J. I. Hage, J. M. Greenberg, and R. T. Wang, "Scattering from arbitrarily shaped particles - theory and experiment," Appl. Opt. **30**, 1141-1152 (1991).

10. F. Rouleau and P. G. Martin, "A new method to calculate the extinction properties of irregularly shaped particles," Astrophys. J. **414**, 803-814 (1993).

11. N. B. Piller, "Influence of the edge meshes on the accuracy of the coupled-dipole approximation," Opt. Lett. **22**, 1674-1676 (1997).

12. N. B. Piller and O. J. F. Martin, "Increasing the performance of the coupled-dipole approximation: A spectral approach," IEEE Trans. Ant. Propag. **46**, 1126-1137 (1998).

13. N. B. Piller, "Coupled-dipole approximation for high permittivity materials," Opt. Comm. **160**, 10-14 (1999).

14. A. G. Hoekstra, J. Rahola, and P. M. A. Sloot, "Accuracy of internal fields in volume integral equation simulations of light scattering," Appl. Opt. **37**, 8482-8497 (1998).

15. S. D. Druger and B. V. Bronk, "Internal and scattered electric fields in the discrete dipole approximation," J. Opt. Soc. Am. B **16**, 2239-2246 (1999).

16. Y. L. Xu and B. A. S. Gustafson, "Comparison between multisphere light-scattering calculations: Rigorous solution and discrete-dipole approximation," Astrophys. J. **513**, 894-909 (1999).

17. M. J. Collinge and B. T. Draine, "Discrete-dipole approximation with polarizabilities that account for both finite wavelength and target geometry," J. Opt. Soc. Am. A **21**, 2023-2028 (2004).

18. M. A. Yurkin, V. P. Maltsev, and A. G. Hoekstra, "Convergence of the discrete dipole approximation. Part 2: an extrapolation technique to increase the accuracy," J. Opt. Soc. Am. A. (to be published).

19. A. Lakhtakia, "Strong and weak forms of the method of moments and the coupled dipole method for scattering of time-harmonic electromagnetic-fields," Int. J. Mod. Phys. C **3**, 583-603 (1992).

20. F. M. Kahnert, "Numerical methods in electromagnetic scattering theory," J. Quant. Spectrosc. Radiat. Transf. **79**, 775-824 (2003).

21. C. P. Davis and K. F. Warnick, "On the physical interpretation of the Sobolev norm in error estimation," Appl. Comp. ElectroMagn. Soc. J. **20**, 144-150 (2005).

22. A. D. Yanghjian, "Electric dyadic Green's function in the source region," IEEE Proc. **68**, 248-263 (1980).

23. G. H. Goedecke and S. G. O'Brien, "Scattering by irregular inhomogeneous particles via the digitized Green's function algorithm," Appl. Opt. **27**, 2431-2438 (1988).

24. C. F. Bohren and D. R. Huffman, *Absorption and scattering of Light by Small Particles*, (Wiley, New York, 1983).

25. J. Rahola, "On the eigenvalues of the volume integral operator of electromagnetic scattering," SIAM J. Sci. Comp. **21**, 1740-1754 (2000).

26. A. Lakhtakia and G. W. Mulholland, "On 2 numerical techniques for light-scattering by dielectric agglomerated structures," J. Res. Nat. Inst. Stand. Technol. **98**, 699-716 (1993).

27. J. I. Hage and J. M. Greenberg, "A model for the optical-properties of porous grains," Astrophys. J. **361**, 251-259 (1990).

28. C. E. Dungey and C. F. Bohren, "Light-scattering by nonspherical particles - a refinement to the coupled-dipole method," J. Opt. Soc. Am. A **8**, 81-87 (1991).

29. H. Okamoto, "Light scattering by clusters: The A1-term method," Opt. Rev. **2**, 407-412 (1995).

30. B. T. Draine and J. J. Goodman, "Beyond clausius-mossotti - wave-propagation on a polarizable point lattice and the discrete dipole approximation," Astrophys. J. **405**, 685-697 (1993).

31. J. I. Peltoniemi, "Variational volume integral equation method for electromagnetic scattering by irregular grains," J. Quant. Spectrosc. Radiat. Transf. **55**, 637-647 (1996).

32. D. Gutkowicz-Krusin and B. T. Draine "Propagation of electromagnetic waves on a rectangular lattice of polarizable points," http://xxx.arxiv.org/abs/astro-ph/0403082 (2004).

33. A. Rahmani, P. C. Chaumet, and G. W. Bryant, "Coupled dipole method with an exact long-wavelength limit and improved accuracy at finite frequencies," Opt. Lett. **27**, 2118-2120 (2002).

34. A. Rahmani, P. C. Chaumet, and G. W. Bryant, "On the importance of local-field corrections for polarizable particles on a finite lattice: Application to the discrete dipole approximation," Astrophys. J. **607**, 873-878 (2004).

35. P. C. Chaumet, A. Sentenac, and A. Rahmani, "Coupled dipole method for scatterers with large permittivity," Phys. Rev. E **70**, 036606 (2004).

36. K. F. Evans and G. L. Stephens, "Microwave radiative-transfer through clouds composed of realistically shaped ice crystals .1. Single scattering properties," J. Atmos. Sci. **52**, 2041-2057 (1995).

37. A. G. Hoekstra, M. D. Grimminck, and P. M. A. Sloot, "Large scale simulations of elastic light scattering by a fast discrete dipole approximation," Int. J. Mod. Phys. C **9**, 87-102 (1998).

38. M. A. Yurkin, K. A. Semyanov, P. A. Tarasov, A. V. Chernyshev, A. G. Hoekstra, and V. P. Maltsev, "Experimental and theoretical study of light scattering by individual mature red blood cells with scanning flow cytometry and discrete dipole approximation," Appl. Opt. **44**, 5249-5256 (2005).

39. "Description of the national compute cluster Lisa," http://www.sara.nl/userinfo/lisa/description/index.html (2005).

## Tables

Table 1. Exact values of $Q_{ext}$ for the 5 test cases.

| particle | $Q_{ext}$ |
|---|---|
| $kD$=8 cube | 4.490 |
| discretized $kD$=10 sphere | 3.916 |
| $kD$=3 sphere | 0.753 |

| $kD$=10 sphere | 3.928 |
|---|---|
| $kD$=30 sphere | 1.985 |

## Figures

1. Fig. 1. Partition of the scatterer's volume into three regions relative to dipole *i*.
2. Fig. 2. $S_{11}(\theta)$ for all 5 test cases in logarithmic scale. The result for the $kD = 3$ sphere is multiplied by 10 for convenience.
3. Fig. 3. Relative errors of $S_{11}$ at different angles θ and maximum over all θ versus *y* for (a) the $kD = 8$ cube, (b) the cubical discretization of $kD = 10$ sphere. A log-log scale is used. A linear fit of maximum over θ errors is shown. ($m = 1.5$).
4. Fig. 4. Same as Fig. 3 but for (a) $kD = 3$, (b) $kD = 10$, and (c) $kD = 30$ spheres.
5. Fig. 5. Relative errors of $Q_{ext}$ versus *y* for all 5 test cases. A log-log scale is used. A linear fit through 5 finest discretizations of $kD = 3$ sphere is shown.

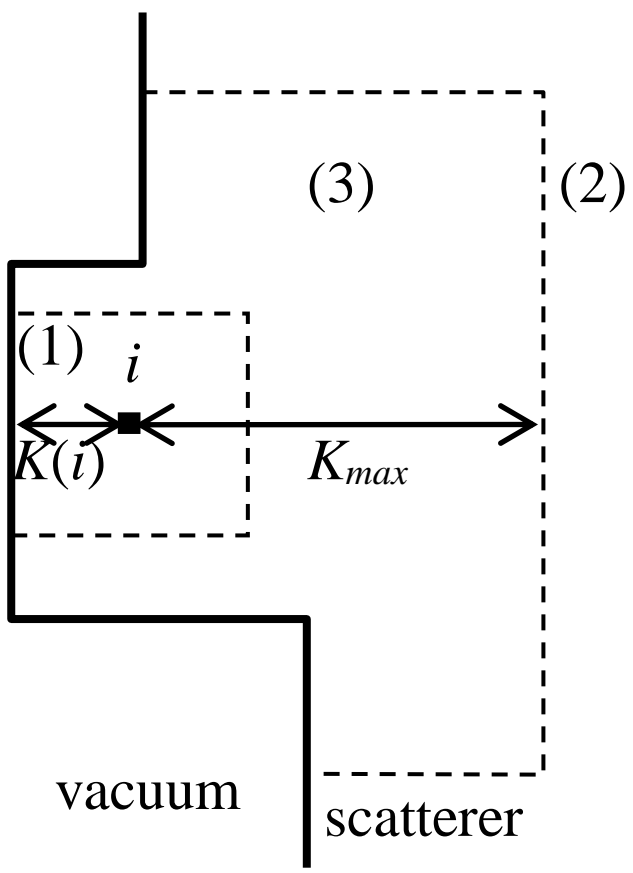

Fig. 1 Yurkin JOSA A

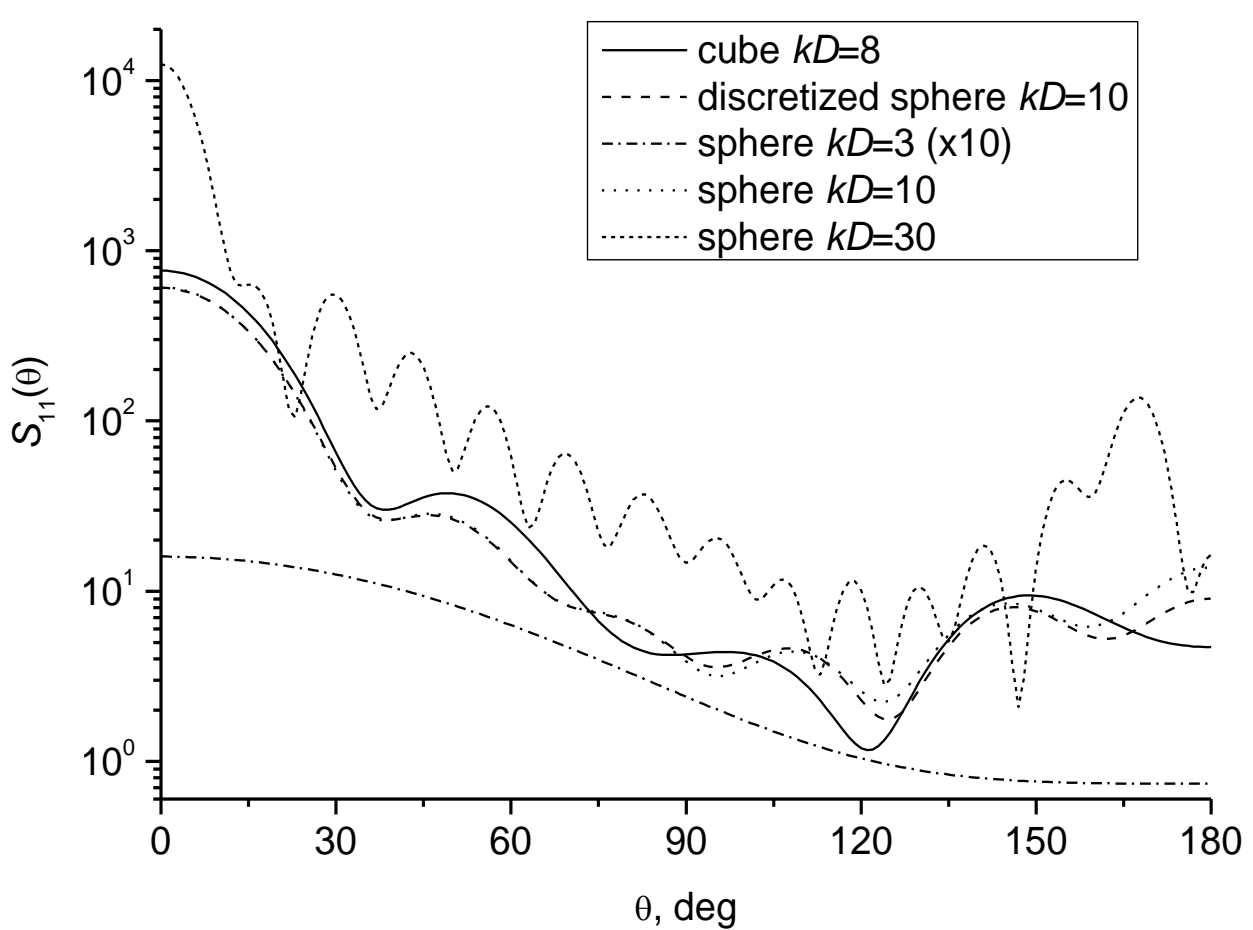

Fig. 2 Yurkin JOSA A

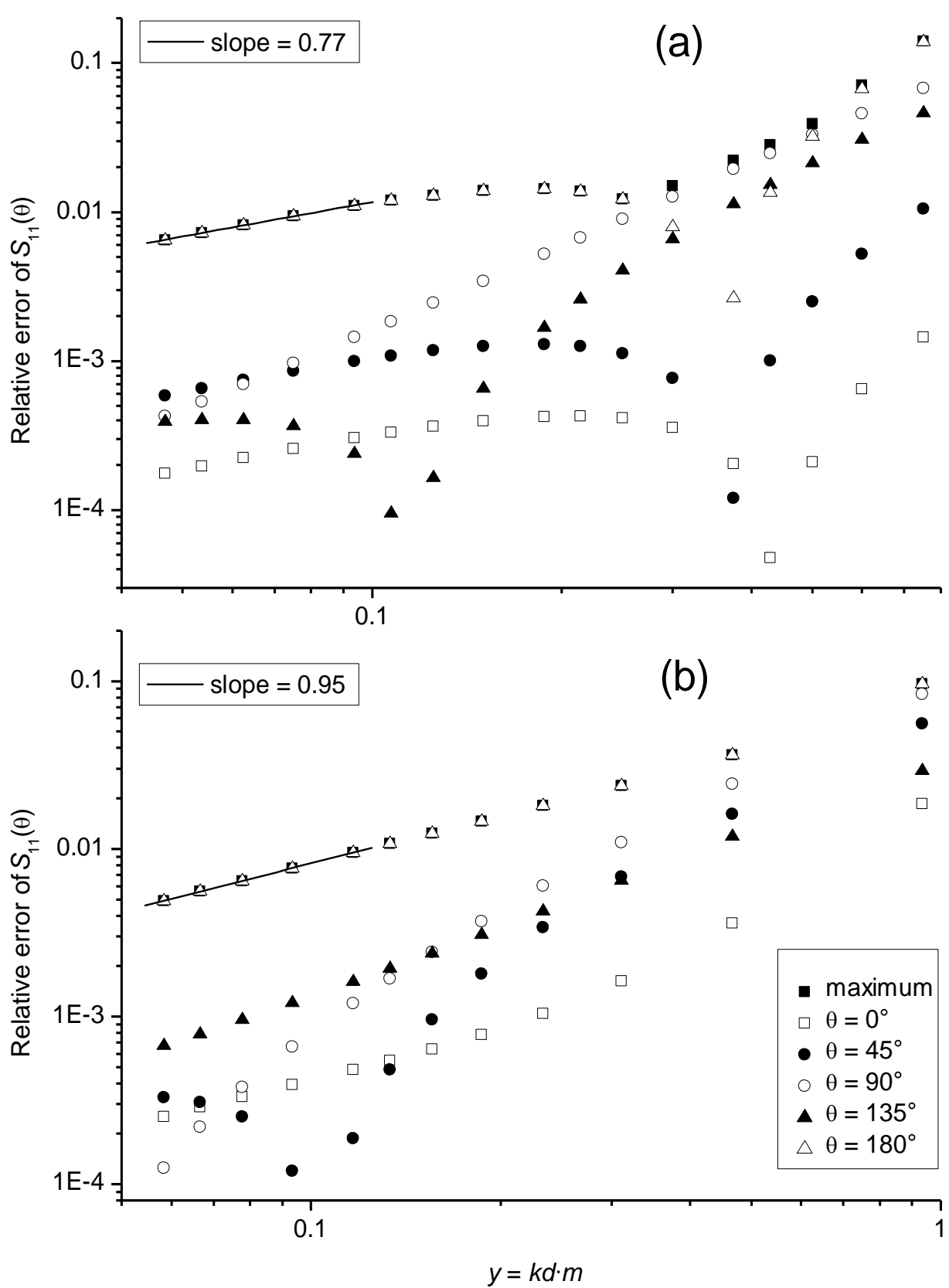

Fig. 3 Yurkin JOSA A

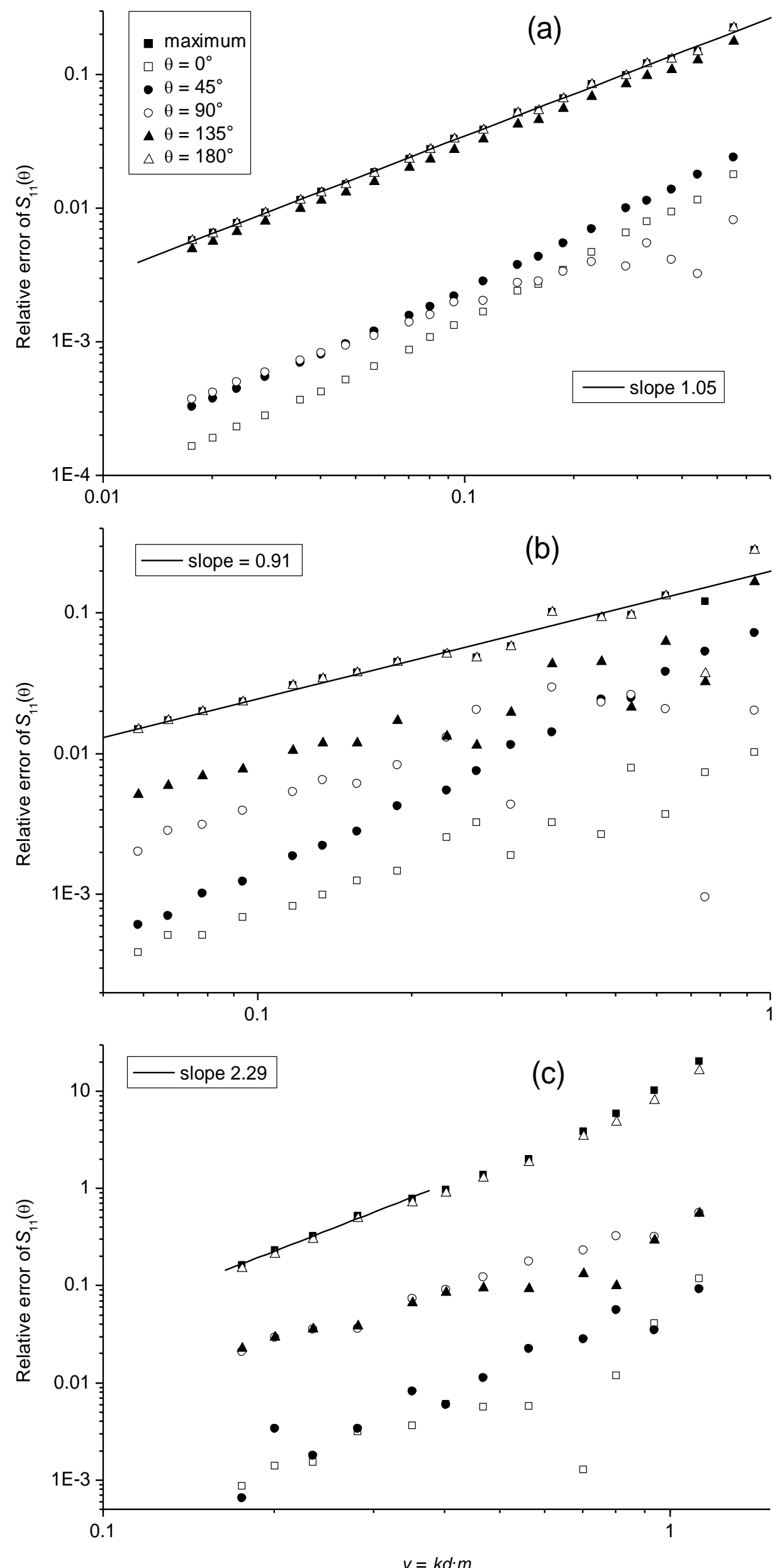

Fig. 4 Yurkin JOSA A

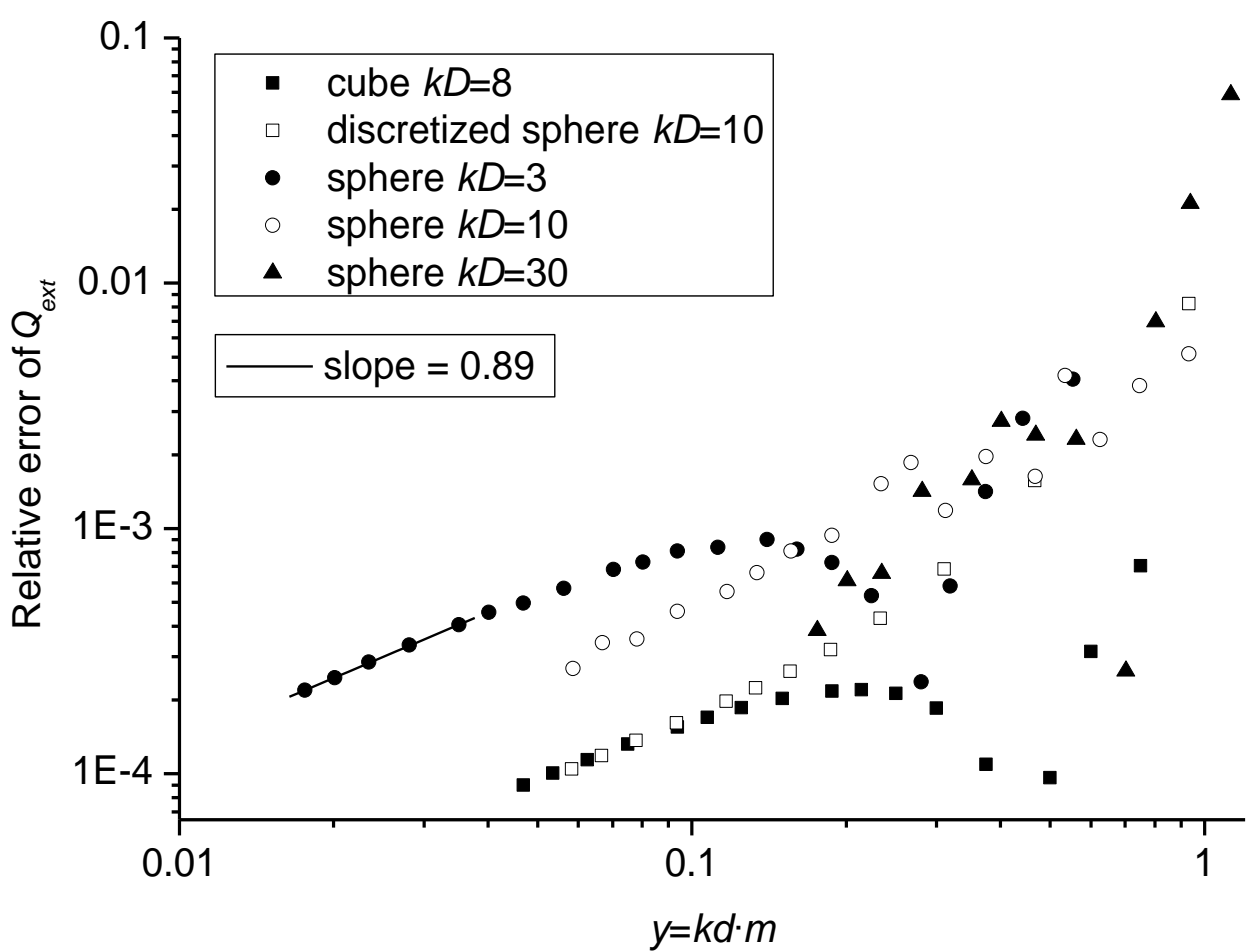

Fig. 5 Yurkin JOSA A